\documentclass[pre, twocolumn, showpacs, preprintnumbers]{revtex4}
\usepackage{amssymb}
\usepackage{amsmath}
\usepackage{graphicx}% Include figure files
\usepackage{dcolumn}% Align table columns on decimal point
\usepackage{bm}% bold math
\usepackage{txfonts}
\DeclareMathAlphabet{\mathpzc}{OT1}{pzc}{m}{it}
\DeclareMathAlphabet{\mathcalligra}{T1}{calligra}{m}{n}

\begin{document}

\preprint{APS/123-QED}

\title{A novel difference between strong liquids and fragile liquids in their dynamics near the glass transition}

\author{Michio Tokuyama$^1$, Shohei Enda$^2$, and Junichi Kawamura$^1$}
\address{$^1$Institute of Multidisciplinary Research for Advanced Materials, Tohoku University, Sendai 980-8577, Japan.\\
$^2$13-36-18, Towada City, Aomori, Japan}

\date{\today}

\begin{abstract}
The systematic method to explore how the dynamics of strong liquids (S) is different from that of fragile liquids (F) near the glass transition is proposed from a unified point of view based on the mean-field theory discussed recently by Tokuyama. The extensive molecular-dynamics simulations are performed on different glass-forming materials. The simulation results for the mean-$n$th displacement $M_n(t)$ are then analyzed from the unified point of view, where $n$ is an even number. Thus, it is first shown that in each type of liquids there exists a master curve $H_n^{(i)}$ as $M_n(t)=R^nH_n^{(i)}(v_{th}t/R;D/Rv_{th})$ onto which any simulation results collapse at the same value of $D/Rv_{th}$, where $R$ is a characteristic length such as an interatomic distance, $D$ a long-time self-diffusion coefficient, $v_{th}$ a thermal velocity, and $i=$F and S. The master curves $H_n^{(F)}$ and $H_n^{(S)}$ are then shown not to coincide with each other in the so-called cage region even at the same value of $D/Rv_{th}$. Thus, it is emphasized that the dynamics of strong liquids is quite different from that of fragile liquids. A new type of strong liquids recently proposed is also tested systematically from this unified point of view. The dynamics of a new type is then shown to be different from that of well-known network glass formers in the cage region, although both liquids are classified as a strong liquid. Thus, it is suggested that a smaller grouping is further needed in strong liquids, depending on whether they have a network or not.
\end{abstract}

\maketitle

\section{Introduction}
It has been known for a long time since Angell \cite{angell} has proposed a famous classification in viscosities of glass-forming materials that there exist two types of glass-forming liquids, fragile liquids (F) and strong liquids (S), near the glass transition \cite{angell91,soko,bohmer93,vil93,angell95,review1,review2,ngai,aip1,ngai11,aip2}. The systems with short-range interactions such as $o$-terphenyl and glycerol are typical examples of fragile liquids, while the covalently bonded network glass formers such as SiO$_2$ and GeO$_2$ are known as typical examples of strong liquids. Thus, it has been understood commonsensically since then that the transport coefficients of both liquids, such as viscosity and self-diffusion coefficient, are well described by the Vogel-Fulcher-Tammann (VFT) law \cite{Vogel,Fulcher,Tammann}, although the fitting temperature range for strong liquids is shorter than that for fragile liquids. However, it is not clear yet how the dynamics of strong liquids is different from that of fragile liquids in a supercooled state. Thus, it is still important to clarify it not only qualitatively but also quantitatively from a unified point of view. 

In order to classify the long-time self-diffusion coefficient $D(T)$ into two types of glass forming liquids from a unified point of view consistently, Tokuyama \cite{toku10,toku11,toku13} has recently shown that the $\alpha$- and $\beta$-relaxation times, $\tau_{\alpha}$ and $\tau_{\beta}$, obey power laws $\tau_{\alpha}\sim D^{-(1+\mu)}$ and $\tau_{\beta}\sim D^{-(1-\mu)}$ in a supercooled state, where the exponent $\mu$ is given by $\mu\simeq 1/5$ for (F) and 2/11 for (S). Then, the following master curve $f(x;\eta)$ for $D(T)$ has been proposed:
\begin{eqnarray}
D(T)&=&d_0f(T_f/T;\eta), \label{1-1}\\
f(x;\eta)&=&\frac{(1-x)^{2+\eta}}{x}\exp[62x^{3+\eta}(1-x)^{2+\eta}], \label{1-2}
\end{eqnarray}
where $T_f$ is a fictive singular temperature to be determined and $d_0$ a positive constant to be determined. Here the exponent $\eta$ is given by $\eta=2(1-3\mu)/3\mu$; $\eta\simeq 4/3$ for (F) and 5/3 for (S). Thus, it has been shown by analyzing many different data that both types of liquids are well described by two types of master curves up to the deviation point $T_n$, below which all the data start to deviate from them and obey the Arrhenius law, where $T_n>T_f$. Here we note that $T_n$ coincides with the so-called thermodynamic glass transition temperature $T_g$ and the master curves can be also fitted by the VFT law well for $T\geq T_n$ \cite{toku11,toku13}. Thus, all the diffusion data in each type collapse onto each single master curve $f(x;\eta)$ (see Fig. \ref{d}). Their material differences are just characterized by a set of parameters ($T_f$, $d_0$, $\eta$). From this viewpoint, therefore, those parameters may correspond to the so-called degree of fragility usually discussed among different systems \cite{angell91,angell95,sco,nov,voi06}. 

One can now distinguish the long-time self-diffusion coefficient of strong liquids from that of fragile liquids safely by using the master curve $f(x;\eta)$. By using such a master curve, we have recently succeeded in creating a new type of strong liquids which is different from usual network glass formers \cite{toku131}. In fact, the static structure factor $S(q)$ of usual network glass formers has the so-called first sharp diffraction peak, which is related to the size of tetrahedron in SiO$_2$ (see Fig. \ref{Sq}). On the other hand, that of a new type does not have such a peak and its structural properties are the same as those of fragile liquids. Thus, there exists another type of strong liquids, that is, non-network glass formers (S$_{non}$), in addition to usual network glass formers (S$_{net}$). Although both types can be classified as strong liquids by using the same master curve $f(x;\eta)$ with $\eta=5/3$, it is not possible to clarify how a new type is different from the usual strong one. This situation is also true for other materials such as Se which has a network structure and is usually believed to be a fragile system. It is interesting to know whether Se is a fragile liquid (F$_{net}$) or not. However, those materials are not investigated here because there is no simulation data available.  

In the present paper, by using the mean-$n$th displacement $M_n(t)(=\langle|\bm{X}_i^{\alpha}(t)-\bm{X}_i^{\alpha}(0)|^n\rangle$), we only investigate the dynamics of glass-forming materials (F$_{non}$), (S$_{net}$), and (S$_{non}$) from a unified point of view based on the mean-field theory \cite {toku06}, where $\bm{X}_i^{\alpha}(t)$ is a position vector of $i$th atom $\alpha$ at time $t$, the brackets an average over the equilibrium ensemble, and $n$ even numbers. Analyses of many data then suggest an existence of a master curve $H_n^{(i)}$ for $M_n(t)$ in each type as
\begin{equation}
M_n(t)=R^nH_n^{(i)}(v_{th}t/R;D/Rv_{th}), \label{1-3}
\end{equation}
where $R$ is the characteristic length such as an interatomic distance, $v_{th}$ the average thermal velocity, and $i$=F$_{non}$, S$_{net}$, and S$_{non}$. Any data in each type are thus shown to collapse onto a single master curve $H_n^{(i)}$ at the same value of $D/Rv_{th}$. Then, we also show that even at the same value of $D/Rv_{th}$ the master curve $H_n^{(i)}$ for type $i$ does not coincide at all with $H_n^{(j)}$ for other type $j(\neq i)$ in the cage region for $\tau_f\leq t\leq \tau_{\beta}$, in which each particle behaves as if it is trapped in a cage mostly formed by neighboring particles, where $\tau_f$ is a mean-free time before which each particle undergoes a ballistic motion. On the other hand, $H_n^{(i)}$ and $H_n^{(j)}$ $(i\neq j)$ are easily shown to coincide with each other both for a short-time region ($t\ll\tau_f$) and for a long-time region ($\tau_{\beta}\ll t$) at the same value of $D/Rv_{th}$. In fact, for both time regions we have 
\begin{equation}
H_n^{(i)}(\tau)=\frac{(n+1)!(3!)^{-n/2}}{(n/2)!}H_2^{(i)}(\tau)^{n/2}\label{1-4-0}
\end{equation}
with
\begin{equation}
H_2^{(i)}(\tau)\simeq\begin{cases}
3\tau^2, & (t\ll \tau_f)\\
6(D/Rv_{th})\tau & (t \gg\tau_{\beta}), 
\end{cases}\label{1-4}
\end{equation}
where $\tau=v_{th}t/R$. Thus, we emphasize that an explicit disagreement in the dynamics of each type appears only in the cage region, although the analytic form of $H_n^{(i)}$ is not known there yet. Finally, we note that although the even number $n$ is taken up to 6 here for simplicity, the same results as those discussed in the present paper also hold for $n\geq 8$.

We begin in Section II by briefly reviewing the mean-field theory recently proposed. We first discuss the mean-field equation for the mean-square displacement and its related characteristic times. Then, we show two types of master curves for the long-time self-diffusion coefficient. One is a master curve for fragile liquids and another is for strong liquids. In Section III, we introduce several potentials to perform extensive molecular-dynamics simulations. In Section IV, we briefly review how physical quantities satisfy the universality near the glass transition. Based on such a universality, we then show that there exist a master curve $H_n^{(i)}$ for the mean-$n$th displacement in each liquid, (F$_{non}$), (S$_{net}$), and (S$_{non}$). In Section V, we show that the master curves $H_n^{(i)}$ and $H_n^{(j)}$ in different types $i$ and $j(\neq i)$ do not coincide with each other in the cage region even at the same value of $D/Rv_{th}$. We conclude in Section VI with a summary.

\section{Mean-field theory}
Here we briefly summarize the mean-field theory of the glass transition (MFT) for molecular systems recently proposed by Tokuyama \cite{toku10,toku11,toku06,toku00,toku07,toku09}.
The mean-field theory consists of the following two essential points:
(A) Mean-field equation for $M_2(t)$ and (B) Two different types of singular functions for $D(T)$, the mean-field curve $g(T_l/T)$ for a liquid state and the master curve $f(T_f/T, \eta)$ for a supercooled state, where $T_l$ and $T_f$ are fictive singular temperatures to be determined and $T_l>T_f$. 

\subsection{Mean-field equation}
The mean-square displacement $M_2(t)$ of $i$th particle $\alpha$ in molecular systems is described by a nonlinear equation \cite{toku06}
\begin{equation}
\frac{d}{dt}M_2(t)=6D+6[v_{th}^2t-D]e^{-M_2(t)/\ell^2},\label{2-1}
\end{equation}
where $\ell$ is a mean-free path of particle $\alpha$ over which the particle can move freely by a ballistic motion and $v_{th}(=(k_BT/m)^{1/2})$ the average thermal velocity. Equation (\ref{2-1}) can be easily solved to give a formal solution
\begin{eqnarray}
&&M_2(t)=6Dt\nonumber\\
&&+\ell^2\ln\left[e^{-6t/\tau_{\beta}}
+\frac{1}{6}\left(\frac{\tau_{\beta}}{\tau_f}\right)^2\left\{1-\left(1+\frac{6t}{\tau_{\beta}}\right)e^{-6t/\tau_{\beta}}\right\}\right],
\label{2-2}
\end{eqnarray}
where $\tau_{\beta}(=\ell^2/D)$ denotes a time for a particle to diffuse over a distance of order $\ell$ with the diffusion coefficient $D$ and is identical to the so-called $\beta$-relaxation time. Here $\tau_f(=\ell/v_{th})$ is a mean-free time, within which each particle can move freely by a ballistic motion. The solution (\ref{2-2}) satisfies the asymptotic forms given by Eq. (\ref{1-4}).
As shown in Ref. \cite{toku07}, the mean-free path $\ell$ is uniquely determined by $D/(Rv_{th})$. Hence the solution (\ref{2-2}) suggests that the dynamics is described by only one parameter $D/(R v_{th})$ if the length and the time are scaled by $R$ and $\tau_{th}(=R/v_{th})$, respectively. This means that the dynamics in different systems coincides with each other if $D/(R v_{th})$ has the same value in them. Hence this is called a universality in dynamics. Since the single-particle dynamics is determined by only one parameter $D/(Rv_{th})$, it is convenient to introduce a new parameter $\hat{u}$ by \cite{toku09}
\begin{equation}
\hat{u}=\log_{10}(Rv_{th}/D).\label{2-9}
\end{equation}
As $\hat{u}$ increases (or $T$ decreases), there exist three states; a liquid state [L] for $\hat{u}<\hat{u}_{\beta}$ (or $T_s<T$), a supercooled state [S] for $\hat{u}_{\beta}\leq \hat{u}<\hat{u}_g$ (or $T_g<T\leq T_s$), and a glass state [G] for $\hat{u}_g\leq\hat{u} $ (or $T\leq T_g$), where $T_s$ is a supercooled point and $T_g$ a glass transition point. The values of $\hat{u}_i$ are listed in Table \ref{table-01}. Here $\hat{u}_{\beta}$ (or $T_s$) is determined by the intersection point of the mean-field curve $g(T_l/T)$ with the master curve $f(T_f/T)$ \cite{toku09} and coincides with a peak position of a specific heat, while $\hat{u}_g$ (or $T_g$) is determined by a deviation point $T_n$ at which the simulation results and the experimental data for the long-time self-diffusion coefficient start to deviate from the master curve $f(T_f/T)$ since $T_n$ coincides with the thermodynamic glass transition point \cite{toku13}. Thus, the mean-field fitting values for the mean-free path $\ell/R$ and the $\beta$-relaxation time $\tau_{\beta}/\tau_{th}$ are uniquely determined by $\hat{u}$. In general, however, the length $R$ is not known. As a well-known example in which $R$ is known, one can take the Lennard-Jones (LJ) binary mixtures A$_{80}$B$_{20}$, where the LJ potential $U_{\alpha\beta}(r)$ is given by
\begin{equation}
U_{\alpha\beta}(r)=4\varepsilon_{\alpha\beta}[(\sigma_{\alpha\beta}/r)^{12}-(\sigma_{\alpha\beta}/r)^6].\label{2-3-1}
\end{equation}
Here $\sigma_{AA}=\sigma$, $\varepsilon_{AA}=\varepsilon$, $\sigma_{AB}=0.8\sigma$, $\varepsilon_{AB}=1.5\varepsilon$, $\sigma_{BB}=0.88\sigma$, and $\varepsilon_{BB}=0.5\varepsilon$, where $\sigma$ is a length unit and $\varepsilon$ an energy unit \cite{kob}. Then, one can choose $\sigma$ as $R$ for A particle. Thus, one can use the simulation results for the LJ binary mixtures as reference to determine $R$ for fragile systems based on the universality. This will be discussed later.
\begin{table}
\caption{Universal value $\hat{u}_i$.}
\begin{center}
\begin{tabular}{cccc}
\hline
type & $\hat{u}_s$ & $\hat{u}_{\beta}$  &$\hat{u}_g$  \\
\hline
fragile & 1.43  & 2.833 & 5.0   \\
strong & 1.5  & 2.693 & 4.0   \\
 \hline
\end{tabular}
\end{center}
\label{table-01}
\end{table}

\subsection{Master curve for long-time self-diffusion coefficient}
In this subsection, we briefly review two types of master curves for $D$.

In order to distinguish the strong liquids from the fragile liquids consistently, Tokuyama has recently analyzed the structural relaxation time $\tau_{\alpha}$ and the $\beta$-relaxation time $\tau_{\beta}$ for self-diffusion in different glass-forming liquids and has proposed two types of master curves for the self-diffusion near the glass transition \cite{toku10,toku11,toku13}. Here $\tau_{\alpha}$ is defined as a time on which the self-intermediate scattering function $F_S(q,t)$ decays to $e^{-1}$ of its initial value, that is, $F_S(q,\tau_{\alpha})=e^{-1}$, while $\tau_{\beta}$ is a time on which the particles can escape from their cages \cite{toku09}. In a liquid state [L], the relaxation times $\tau_{\alpha}$ and $\tau_{\beta}$ are then shown to obey power laws
\begin{equation}
\tau_{\alpha}\sim\tau_{\beta}\sim D^{-(1-\nu)}, \label{tau0}
\end{equation}
where the exponent $\nu$ is obtained by fitting as $\nu\simeq 1/3$. In [L], the experimental data and the simulation results can be well described by the mean-field singular function \cite{toku09,toku091}
\begin{equation}
D(T)\propto g(T_l/T) \propto\left(\frac{T}{T_l}\right)\left(1-\frac{T_l}{T}\right)^2 \sim \epsilon_0^2, \label{dif0}
\end{equation}
where $T_l$ is a singular temperature to be determined by fitting and $\epsilon_0=1-T_l/T$.
On the other hand, in a supercooled state [S], the relaxation times $\tau_{\alpha}$ and $\tau_{\beta}$ are shown to obey power laws
\begin{equation}
\tau_{\alpha}\sim D^{-(1+\mu)}, \;\;\;\tau_{\beta}\sim D^{-(1-\mu)}, \label{tau}
\end{equation}
where the exponent $\mu$ is obtained by fitting as $\mu\simeq 1/5$ for fragile liquids and 2/11 for strong liquids. We now assume that as long as the system is in equilibrium, the long-time self-diffusion coefficient $D(T)$ obeys the following singular function in [S]:
\begin{equation}
D(T)\propto \left(\frac{T}{T_f}\right)\left(1-\frac{T_f}{T}\right)^{2+\eta} \sim \epsilon^{2+\eta}, \label{dif1}
\end{equation}
where $T_f (<T_l)$ is a new fictive singular temperature to be determined by fitting and $\epsilon=1-T_f/T$. Here the exponent $\eta$ is obtained as follows. The $\beta$-relaxation time $\tau_{\beta}$ is given by $\tau_{\beta}=\ell^2/D$. Then, use of Eqs. (\ref{tau0})-(\ref{dif1}) leads to
\begin{eqnarray}
\ell &\sim& \epsilon_0^{\nu}  \;\;\;\;\;\;\;\;\;\;\;\;\text{in [L]}, \label{L}\\
\ell &\sim& \epsilon^{(2+\eta)\mu/2} \;\;\;\;\text{in [S]}. \label{S} 
\end{eqnarray}
The detailed analyses \cite{toku091,toku07} show that $\ell$ obeys the same power law in both states because the caging mechanism does not change in both states. Thus, use of Eqs. (\ref{L}) and (\ref{S}) leads to
\begin{equation}
\nu=(2+\eta)\mu/2 \;\;\;   \text{or}\;\;\;   \eta=2(\nu/\mu-1).\label{eta}
\end{equation}
Then, one finds $\eta\simeq 4/3$ for fragile liquids and 5/3 for strong liquids. Thus, Eq. (\ref{dif1}) can describe the self-diffusion data in a supercooled state, while Eq. (\ref{dif0}) holds in a liquid state. The intersection point of Eq. (\ref{dif1}) with Eq. (\ref{dif0}) thus determines a supercooled point $T_s$ (or $u_{\beta}$). In order to find an asymptotic function which holds in both states, we assume that $D(T)$ can be written as
\begin{equation}
D(T)=d_0f(T_f/T; \eta), \label{mc0}
\end{equation} 
where $d_0$ is a positive constant to be determined. Then, the function $f(x)$ must numerically coincide with Eq. (\ref{dif0}) in [L] and Eq. (\ref{dif1}) in [S]. As shown in Ref. \cite{toku11}, expanding $f(x)$ in powers of $\epsilon^{2+\eta}/x$, one can thus find the master curve
\begin{equation}
f(x; \eta)\simeq \frac{(1-x)^{2+\eta}}{x}\exp[62x^{3+\eta}(1-x)^{2+\eta}], \label{MC}
\end{equation}
where $x=T_f/T$. We note here that the power-law exponent for strong liquids is slightly different from that for fragile liquids. Although the quantitative difference between exponents in both liquids is small, it is important to show that there exist qualitatively different mechanisms between them since the exponents should result from the many-body correlations. In the previous paper \cite{toku11}, we have shown that Eq. (\ref{MC}) can describe any data for self-diffusion coefficient in fragile and strong liquids, up to the deviation temperature $T_n$, below which the system becomes out of equilibrium and all the data start to deviate from the master curve. If one scales the data by $d_0$ and $T_f$, therefore, they are all collapsed onto two types of master curves given by Eq. (\ref{MC}), a fragile master curve with $\eta=4/3$ and a strong master curve with $\eta=5/3$ (see Fig. \ref{d}). Thus, one can classify the microscopic differences among various liquids by a set of parameters ($T_f$, $d_0$, $\eta$). Hence it must give another expression for the so-called fragility of glass-forming materials.

In the previous papers \cite{toku10,toku11,toku13}, we have investigated many different glass-forming materials from a unified point of view based on two types of master curves $f(x;\eta)$ and classified them into two types of liquids, fragile liquids with $\eta=4/3$ and strong liquids with $\eta=5/3$. However, this classification has two weak points. The first is that the difference between their exponents $\eta$ is very small to distinguish two types of liquids quantitatively. Hence very precise analyses are required to make it even under the situation that the experimental data and the simulation results always have fluctuations. The second is that as discussed before, it can not distinguish non-network glass formers from network glass formers because they have the same value of $\eta$ as 5/3. In the following, therefore, we first perform the extensive molecular-dynamics simulations on different glass-forming materials and then investigate their dynamics fully by calculating the mean-$n$th displacement $M_n(t)$. Thus, we show how the dynamics of strong liquids is different from that of fragile liquids consistently from a unified point of view based on the universality.

\section{Molecular-dynamics simulations}
In order to investigate the differences between fragile liquids and strong liquids, we perform the extensive molecular-dynamics simulations under the so-called $NVT$ method with periodic boundary conditions on the following different systems: For fragile liquids we take binary mixtures A$_{80}$B$_{20}$ with the Stillinger-Weber (SW) potential \cite{sw} and Al$_2$O$_3$ with the Born-Meyer (BM) potential \cite{bm}. We also use the previous simulation results on the LJ binary mixtures \cite{narumi07,narumi11} as reference. On the other hand, for strong liquids we take SiO$_2$ with the Beest-Kremer-Santen (BKS) potential \cite{bks} and also SiO$_2$ with the Nakano-Vashishta (NV) potential \cite{nv} as a typical example of network glass formers. We also consider A$_{80}$B$_{20}$ with the SW potential under different mass ratios as a typical example of non-network glass formers \cite{toku131}.

The SW potential is given by
\begin{equation}
U_{\alpha\beta}(r)=\begin{cases}
\varepsilon_{\alpha\beta}[\left(\frac{\sigma_{\alpha\beta}}{r}\right)^{12}-1]\exp\left[\left(\frac{r}{\sigma_{\alpha\beta}}-r_c\right)^{-1}\right]\;\; \text{for}\;\; \frac{r}{\sigma_{\alpha\beta}}<r_c,\\
0 \;\;\text{for}\;\; \frac{r}{\sigma_{\alpha\beta}}>r_c,
\end{cases}\label{3-1}
\end{equation}
where $\alpha, \beta \in \{A,B\}$. Here the parameters $\varepsilon_{\alpha\beta}$, $\sigma_{\alpha\beta}$, and $r_c$ are given by $\sigma_{AA}=\sigma$, $\varepsilon_{AA}=8.8\varepsilon$, $\sigma_{AB}=0.8\sigma$, $\varepsilon_{AB}=13.2\varepsilon$, $\sigma_{BB}=0.88\sigma$, $\varepsilon_{BB}=4.4\varepsilon$, and $r_c=1.652$. Here $\varepsilon$ is an energy unit and $\sigma$ a length unit. The system contains $N=10976$ particles, which is composed of $N_A=8780$ particles of type A with mass $m_A$ and $N_B=2196$ particles of type B with mass $m_B$. Length, time, and temperature are scaled by $\sigma$, $t_0 (=\sigma/v_0)$, and $\varepsilon/k_B$, respectively, where $v_0 = (\varepsilon/m_A)^{1/2}$. The simulations are performed in a cubic box of length 20.89$\sigma$ with periodic boundary conditions, where the number density is 1.2. As shown in the previous paper \cite{toku131}, for $Q(=m_B/m_A)<Q_c$ the system shows dynamic properties of fragile liquids, while for $Q>Q_c$ the system shows those of strong liquids, where $Q_c\simeq 20$. Here we note that their static structure factors do not depend on $Q$. Hence we call those strong liquids non-network glass formers, distinguishing from usual network glass formers, such as SiO$_2$.

The BKS and the BM potentials are given by
\begin{equation}
U_{\alpha\beta}(r)=\frac{q_{\alpha}q_{\beta}}{r_{\alpha\beta}}+A_{\alpha\beta}\exp(-b_{\alpha\beta}r_{\alpha\beta})-\frac{c_{\alpha\beta}}{r_{\alpha\beta}^6},\label{3-2}
\end{equation}
where the potential parameters are listed in Table \ref{table-2}.
For the BKS potential, the system contains $N=3000$ particles in the cubic box of volume $L^3$, which is composed of $N_{Si}=1000$ particles of Si with mass $m_{Si}=4.66\times 10^{-26}$ (kg) and $N_O=2000$ particles of O with $m_O=2.66\times 10^{-26}$ (kg), where $L=34.79\AA$.  For the BM potential, the system contains $N=3000$ particles in the cubic box of volume $L^3$, which is composed of $N_{Al}=1200$ particles of Al with mass $m_{Al}=4.5\times 10^{-26}$ (kg) and $N_O=1800$ particles of O with $m_O=2.66\times 10^{-26}$ (kg), where $L=32.02\AA$. Those system sizes are enough to avoid a finite size effect in strong liquids \cite{hor}.
\begin{table}
\caption{Potential parameters for BKS \cite{bks} and BM \cite{bm}.}
\begin{center}
\begin{tabular}{cccccc}
\hline
& & $q_{\alpha}$ (e)  & $A_{\alpha\beta}$ (eV) &$b_{\alpha\beta}$($\AA^{-1}$) &$c_{\alpha\beta}$ (eV$\AA^6)$  \\
\hline
&Si-Si& 2.4 & 0.0000 & 0.00000  & 0.0000  \\
BKS&O-Si& - & 18003.7572 &4.87318  & 133.5381 \\
&O-O & -1.2 & 1388.7730 & 2.76000 & 175.0000\\
\hline
&Al-Al & 3 & 0.00 & 3.448 & 0.0000\\
BM&Al-O & - & 1779.86 & 3.448 & 0.0000\\
&O-O & -2 & 1500.00 & 3.448 &0.0000\\
\hline
\end{tabular}
\end{center}
\label{table-2}
\end{table}

The NV potential is given by
\begin{equation}
U=\sum_{\alpha<\beta}U^{(2)}_{\alpha\beta}+\sum_{\alpha, \beta<\gamma}U^{(3)}_{\alpha\beta\gamma}\label{3-3}
\end{equation}
with the two-body potential
\begin{eqnarray}
U^{(2)}_{\alpha\beta}(r)&=&\epsilon\left(\frac{a_{\alpha\alpha}+a_{\beta\beta}}{r}\right)^{n_{\alpha\beta}}+\frac{Z_{\alpha}Z_{\beta}}{r}e^{-r/A_0}\nonumber\\
&-&\frac{a_{\alpha}Z_{\beta}^2+a_{\beta}Z_{\alpha}^2}{2r^4}e^{-r/A_1},\label{3-4}
\end{eqnarray}
and the three-body potential
\begin{eqnarray}
U^{(3)}_{\alpha\beta\gamma}&=&
B_{\alpha}\exp\left[\frac{1}{r_{\alpha\beta}-A_2}+\frac{1}{r_{\alpha\gamma}-A_2}\right]\nonumber\\
&\times&\left(\frac{\bm{r}_{\alpha\beta}\cdot\bm{r}_{\alpha\gamma}}{r_{\alpha\beta}r_{\alpha\gamma}}-\cos\overline{\theta}_{\alpha}\right)^2\theta(A_2-r_{\alpha\beta})\theta(A_2-r_{\alpha\gamma}),\nonumber\\
\label{3-5}
\end{eqnarray}
where $\theta(x)$ is a step function, $A_0=4.43 (\AA)$, $A_1= 2.5 (\AA)$, $A_2=5.5 (\AA)$. The potential parameters are listed in Table \ref{table-3}.
\begin{table}
\caption{Potential parameters for NV \cite{nv}.}
\begin{center}
\begin{tabular}{cccccccc}
\hline
 & $\epsilon$ (eV)  & $a_{\alpha\alpha} (\AA)$ &$Z_{\alpha}$(e) &$a_{\alpha} (\AA^3)$ & $n_{\alpha\beta}$&$B_{\alpha}$ (eV)&$\overline{\theta}_{\alpha}$ \\
\hline
Si-Si & 1.592 & 0.47 & -0.88  & 0.00  &11&-&-\\
O-Si & 1.592 & - & - & -  &9&-&-\\
O-O& 1.592 & 1.2 & 1.76  & 2.4  &7&-&-\\
O-Si-O &-&-&-&-&-&4.993&109.47\\
Si-O-Si &-&-&-&-&-&19.972&141.00\\
\hline
\end{tabular}
\end{center}
\label{table-3}
\end{table}
The system contains $N=5184$ particles in the cubic box of volume $L^3$, which is composed of $N_{Si}=1728$ particles of Si with mass $m_{Si}=4.66\times 10^{-26}$ (kg) and $N_O=3456$ particles of O with $m_O=2.66\times 10^{-26}$ (kg), where $L=42.996\AA$. 

The Newton equations are solved by the velocity Verlet algorithm under the NVT ensemble for each system. The simulations are repeated until the system is equilibrated, where the time scale of equilibration is of order 10 ns for BKS, BM, and NV, and of order $10^5t_0$ for SW. Next we analyze those simulation results from a unified point of view based on the universality.

\section{Universality near the glass transition}
We now analyze the simulation results in two different types of liquids, fragile liquids and strong liquids, from a unified point of view based on the universality and then show how the dynamics of strong liquids is different from that of fragile liquids.

In order to discuss the differences between strong liquids and fragile liquids, we investigate the following physical quantities. The first is the mean-$n$th displacement given by
\begin{equation}
M_n(t)=<|\bm{X}_i^{\alpha}(t)-\bm{X}_i^{\alpha}(0)|^n>,\label{4-1}
\end{equation}
where the brackets indicate the average over an equilibrium ensemble, $\bm{X}_i^{\alpha}(t)$ the position vector of $i$th particle $\alpha$ at time $t$, and $n$ even numbers. The second is the long-time self-diffusion coefficient $D(T)$ given by
\begin{equation}
D(T)=\lim_{t\rightarrow \infty}\frac{M_2(t)}{6t}.\label{4-2}
\end{equation}
The third is the static structure factor $S_{\alpha\beta}(q)$ given by
\begin{equation}
S_{\alpha\beta}(q)=\frac{1}{N}\sum_{i=1}\sum_{j\neq i}<\exp[i\bm{q}\cdot\{\bm{X}_i^{\alpha}(0)-\bm{X}_j^{\beta}(0)\}]>.\label{4-3}
\end{equation}

\begin{figure}
\begin{center}
\includegraphics[width=8.0cm]{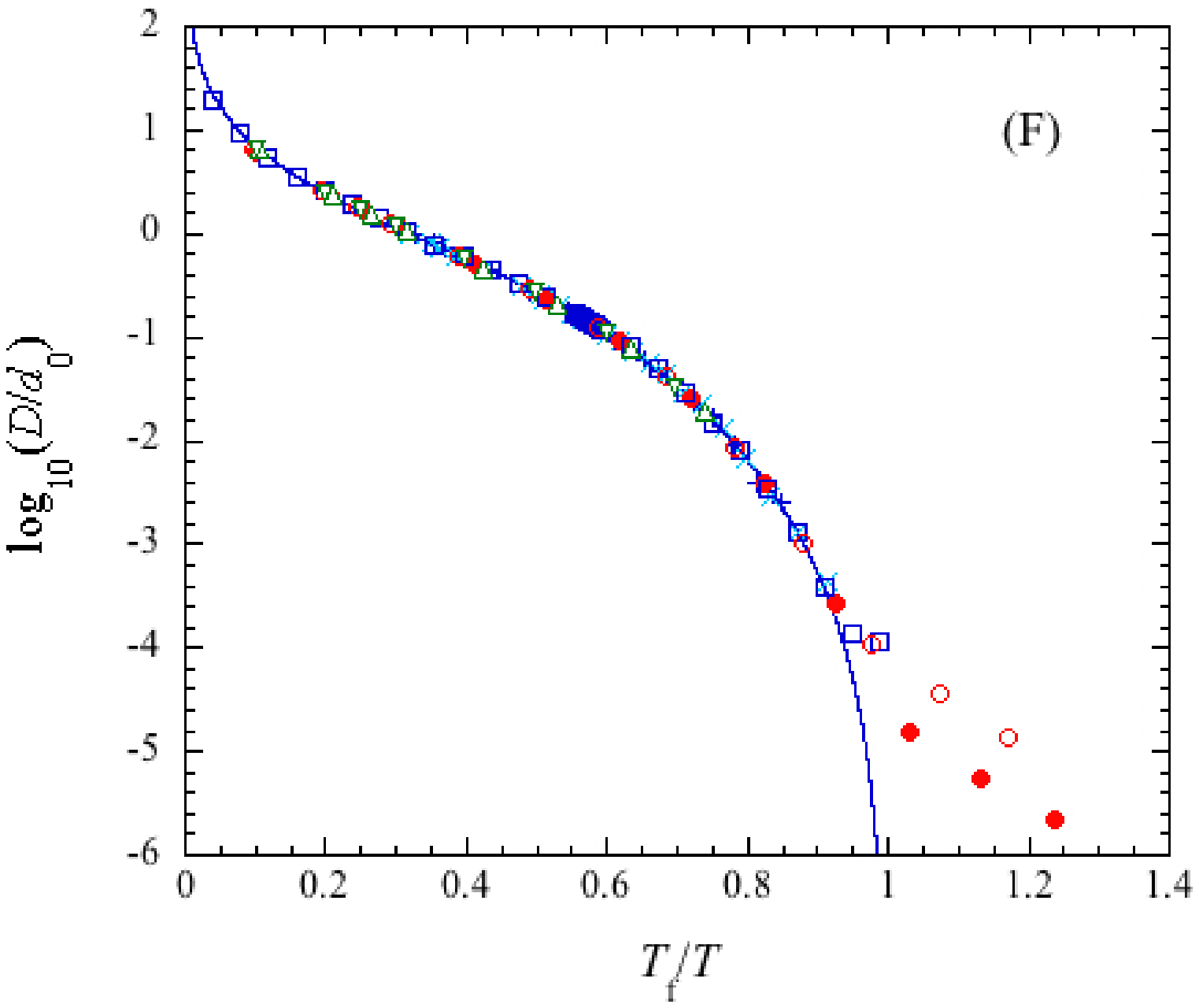}\hspace{0.0cm}
\includegraphics[width=8.0cm]{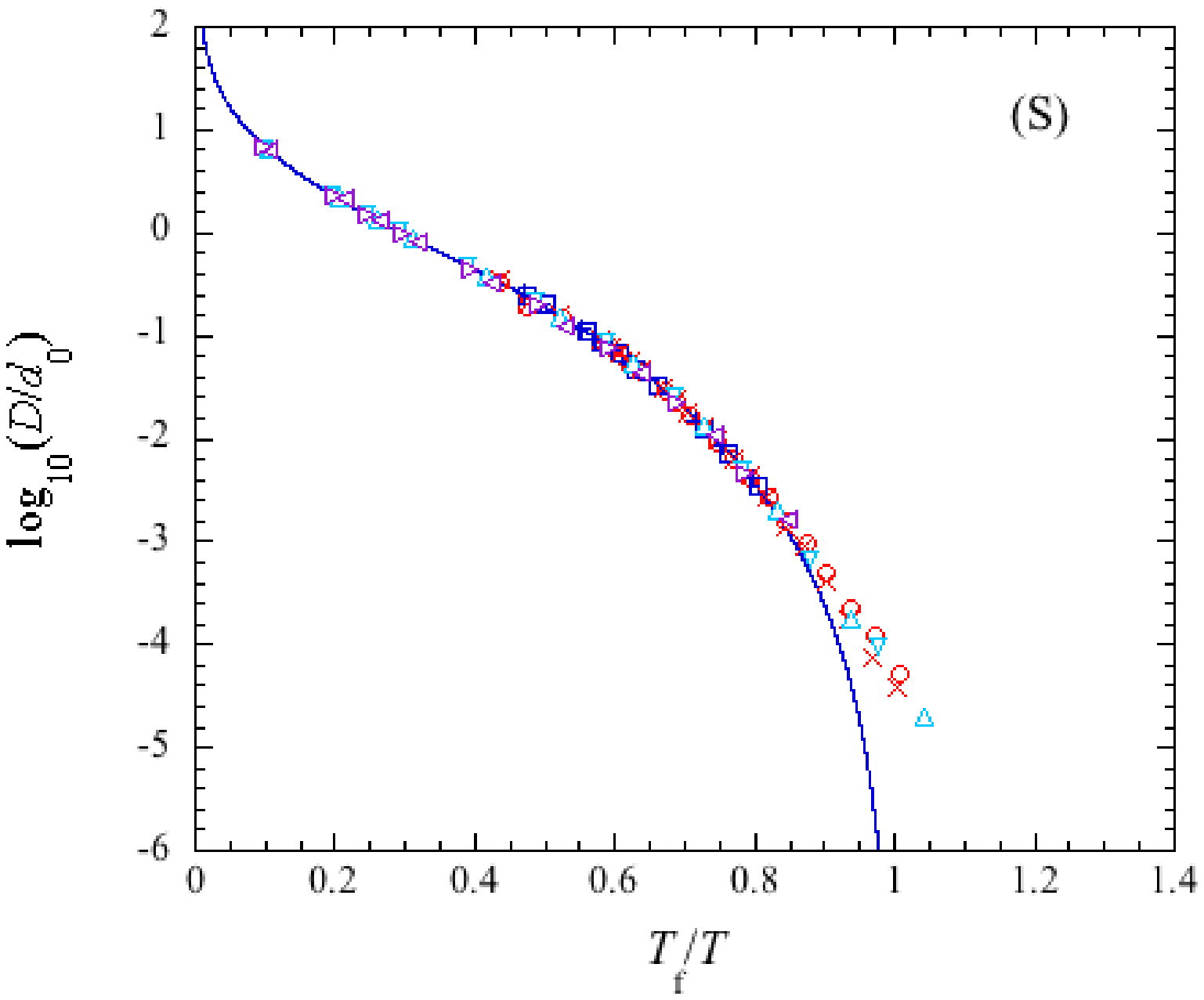}\hspace{0.0cm}
\end{center}
\caption{(Color online) A log-log plot of $D/d_0$ versus scaled temperature $T_f/T$ for fragile liquids (F) and strong liquids (S). The  symbols indicate the simulation results; for (F) $(\Box)$ A in A$_{80}$B$_{20}$ (LJ), $(+)$ Al in Al$_2$O$_3$, $(\times)$ O in Al$_2$O$_3$,  $(\bullet)$ A in A$_{80}$B$_{20}$ (SW) with $Q=1$, $(\circ)$ B in A$_{80}$B$_{20}$ (SW) with $Q=1$, $(\triangle)$ A in A$_{80}$B$_{20}$ (SW) with $Q=10$, and $(\triangledown)$ B in A$_{80}$B$_{20}$ (SW) with $Q=10$, and for (S) $(\Box)$ Si in SiO$_2$ (BKS), $(+)$ O in SiO$_2$ (BKS), $(\times)$ Si in SiO$_2$ (NV), $(\circ)$ O in SiO$_2$ (NV), $(\triangle)$ A in A$_{80}$B$_{20}$ (SW) with $Q=50$, $(\triangledown)$ B in A$_{80}$B$_{20}$ (SW) with $Q=50$, $(\lhd)$ A in A$_{80}$B$_{20}$ (SW) with $Q=100$, and $(\rhd)$ B in A$_{80}$B$_{20}$ (SW) with $Q=100$. The solid line indicates the master curve $f(x)$ given by Eq. (\ref{MC}). The relevant parameters $T_f$ and $d_0$ are listed in Table \ref{table-4}.}
\label{d}
\end{figure}
By using (\ref{4-2}), one can first obtain the temperature dependence of $D(T)$ for each system. In Fig. \ref{d}, all the simulation results for $D(T)$ are shown to collapse onto two types of master curves given by Eq. (\ref{MC}), where $T_f$ and $d_0$ are listed in Table \ref{table-4}. As discussed in the previous paper \cite{toku13}, the deviation point $T_n$ at which all data start to deviate from the master curve coincides with the thermodynamic glass transition point $T_g$. For fragile liquids it is given by $T_g\simeq T_f/0.938$ at $f(T_f/T_g)\simeq 10^{-4.0}$, while $T_g\simeq T_f/0.855$ at $f(T_f/T_g)\simeq 10^{-3.0}$ for strong liquids.
\begin{table}
\caption{Characteristic length $R$, singular temperature $T_f$, and $d_0$ for different systems.}
\begin{center}
\begin{tabular}{ccccc}
\hline
type & system & $R$ & $T_f$ & $d_0$\\
\hline
fragile &A (LJ)&1.000 & 0.3955 & 0.0283\\
($\eta=$4/3)& Al (BM) & 2.928($\AA$) & 1955(K) & 0.8394$\times 10^{-8}$(m$^2$/s) \\
& O (BM) & 3.047($\AA$) & 1916(K) & 1.3405$\times 10^{-8}$(m$^2$/s)\\
&A (SW $Q=1$)&1.000 &0.5153 & 0.0267\\
&B (SW $Q=1$) &1.250 &0.4886 & 0.0357\\
&A (SW $Q=10$)&1.000 & 0.5279& 0.0208\\
&B (SW $Q=10$) &1.400 & 0.4983& 0.0236\\
\hline
strong &Si (NV) & 3.717($\AA$) &2612(K) & 4.4367$\times 10^{-8}$(m$^2$/s)\\
($\eta=$5/3)&O (NV) &4.506($\AA$) & 2621(K) &5.7416$\times 10^{-8}$(m$^2$/s)\\
&Si (BKS) & 3.840($\AA$) & 2904(K) & 5.4967$\times 10^{-8}$(m$^2$/s)\\
& O (BKS) & 4.650($\AA$) & 2865(K) &6.5068$\times 10^{-8}$(m$^2$/s)\\
&A (SW $Q=50$)&1.000 & 0.5204 & 0.0183\\
&B (SW $Q=50$) &1.540 & 0.4881& 0.0176\\
&A (SW $Q=100$)&1.000 & 0.5307 & 0.0171\\
&B (SW $Q=100$) &1.600 & 0.4920 & 0.0147\\
\hline
\end{tabular}
\end{center}
\label{table-4}
\end{table}

\begin{figure}
\begin{center}
\includegraphics[width=8.0cm]{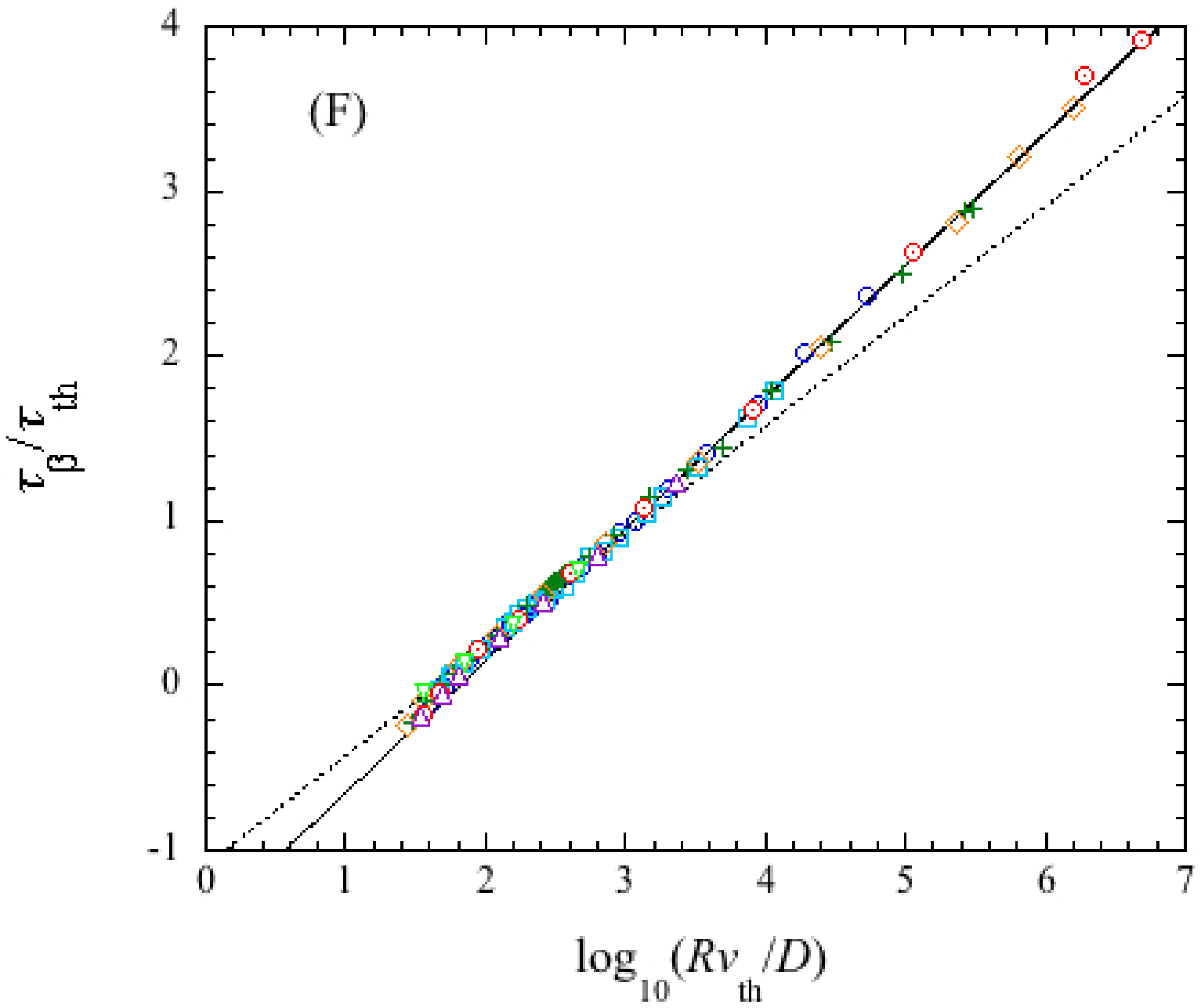}\hspace{0.0cm}
\includegraphics[width=8.0cm]{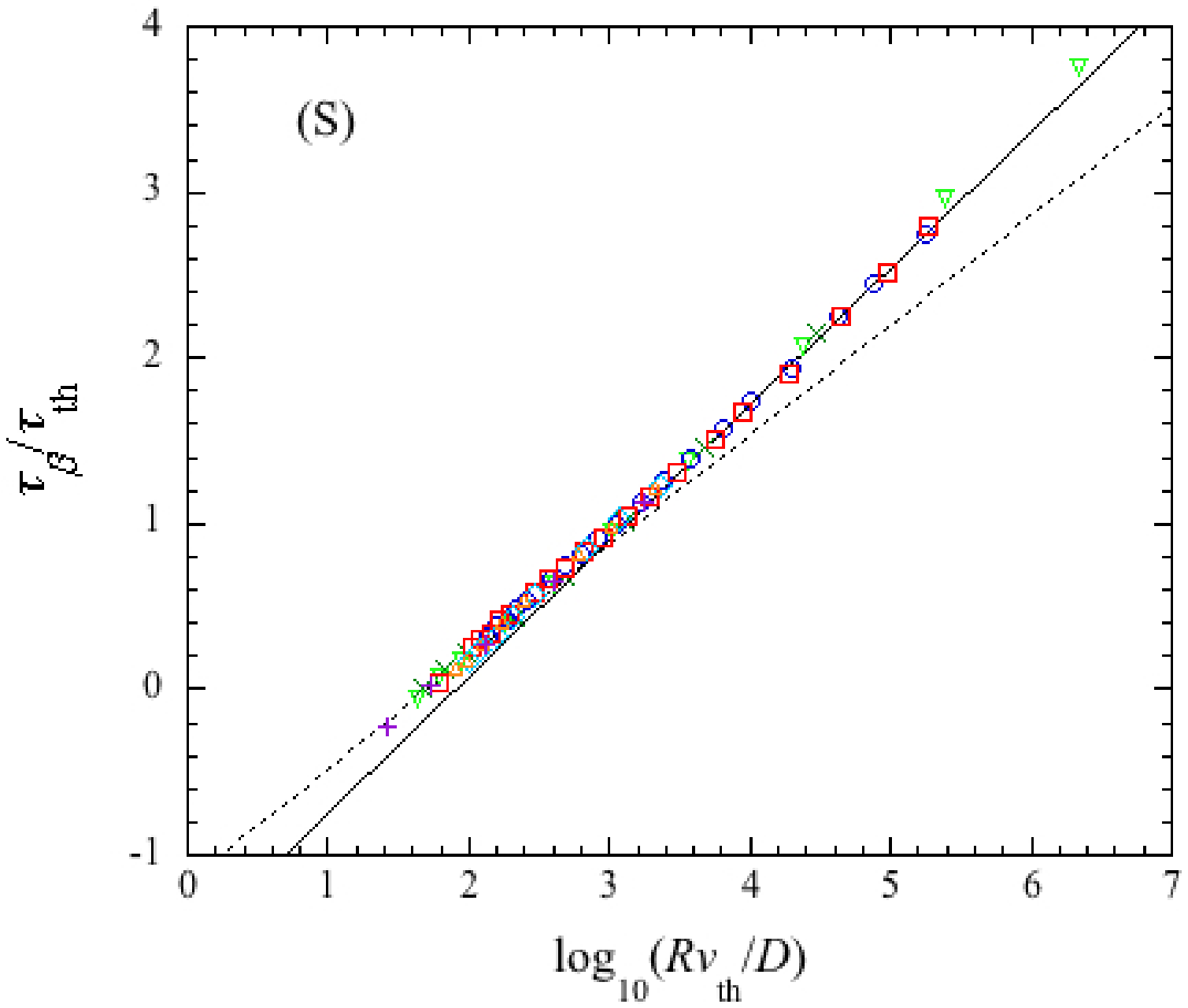}\hspace{0.0cm}
\end{center}
\caption{(Color online) A log-log plot of $\tau_{\beta}/\tau_{th}$ versus $\hat{u}$ for fragile liquids (F) and strong liquids (S). The symbols indicate the simulation results; for (F) $(\Box)$ Al in Al$_2$O$_3$, $(\circ)$ O in Al$_2$O$_3$, $(+)$ A in A$_{80}$B$_{20}$ (LJ), $(\odot)$ A in A$_{80}$B$_{20}$ (SW) with $Q=1$, $(\Diamond)$ B in A$_{80}$B$_{20}$ (SW) with $Q=1$, $(\triangle)$ A in A$_{80}$B$_{20}$ (SW) with $Q=10$, and $(\triangledown)$ B in A$_{80}$B$_{20}$ (SW) with $Q=10$, and for (S) $(\Box)$ Si in SiO$_2$ (NV), $(\circ)$ O in SiO$_2$ (NV), $(\triangle)$ Si in SiO$_2$ (BKS), $(\Diamond)$ O in SiO$_2$ (BKS), $(\times)$ A in A$_{80}$B$_{20}$ (SW) with $Q=100$, $(+)$ B in A$_{80}$B$_{20}$ (SW) with $Q=100$, and $(\triangledown)$ A in A$_{80}$B$_{20}$ (SW) with $Q=50$. The solid lines indicate straight lines $y=0.8x-1.45$ for (F) and $y=(9/11)x-1.55$ for (S), and the dotted lines $y=(2/3)x-1.091$ for (F) and $y=(2/3)x-1.142$ for (S). }
\label{ti}
\end{figure}
\begin{figure}
\begin{center}
\includegraphics[width=8.0cm]{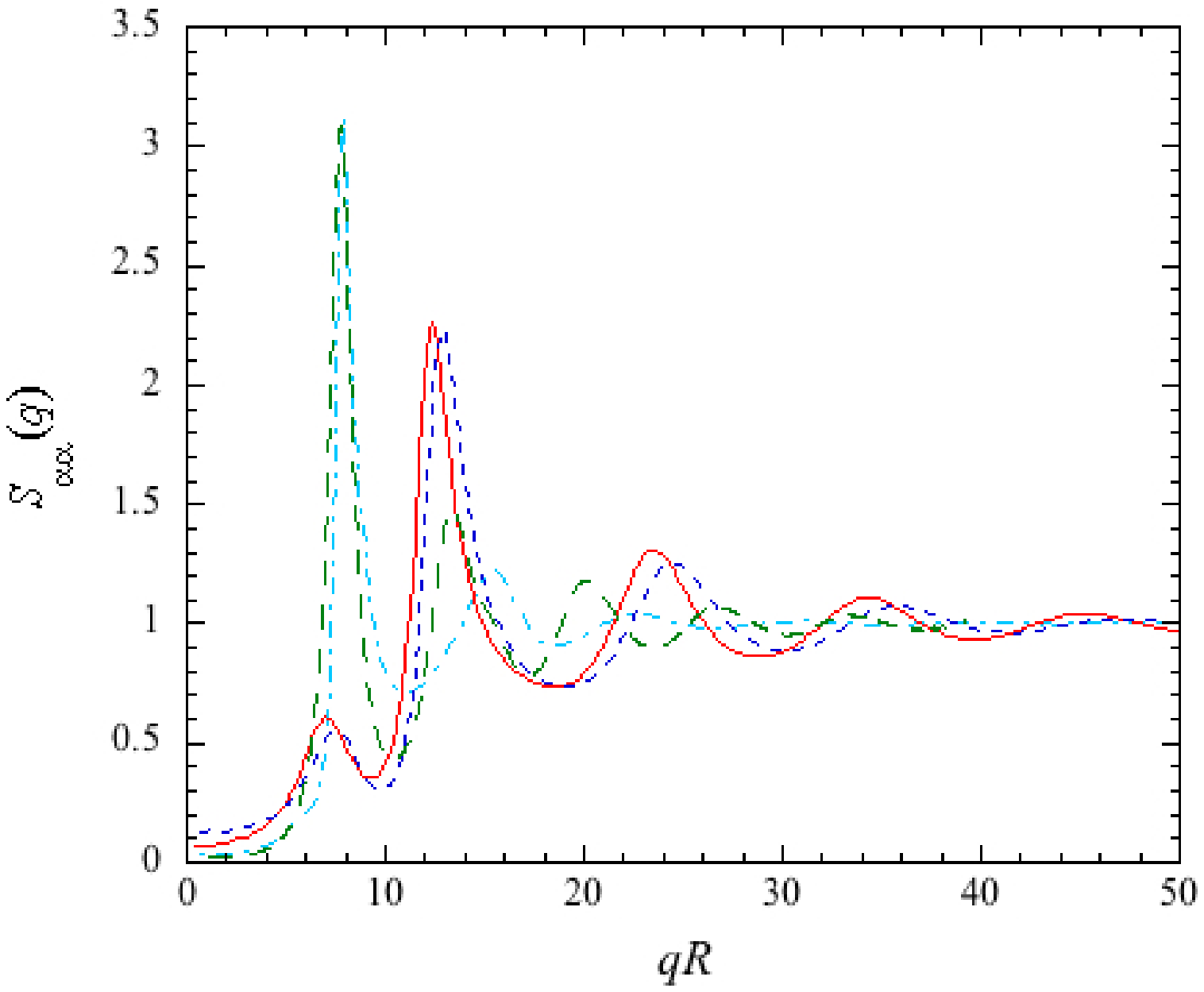}
\end{center}
\caption{(Color online) A plot of $S_{OO}(q)$ and $S_{AA}(q)$ versus scaled wave vector $qR$ around $\hat{u}\simeq 3.143$ for fragile liquids and strong liquids. The solid line indicates $S_{OO}(q)$ for SiO$_2$ (NV) at $T=3500$(K), the dashed line $S_{OO}(q)$ for SiO$_2$ (BKS) at $T=3700$(K),  the dot-dashed line $S_{OO}(q)$ for Al$_2$O$_3$ at $T=2700$ (K), and the long-dashed line $S_{AA}(q)$ for A$_{80}$B$_{20}$ (SW $Q=$1 and 100) at $T$=0.714.}
\label{Sq}
\end{figure}
By using Eq. (\ref{2-2}), one can next obtain the mean-field fitting values for the $\beta$-relaxation time $\tau_{\beta}$. From a unified point of view based on the universality discussed in the previous paper \cite{toku09}, the dimensionless time $\tau_{\beta}/\tau_{th}$ for different systems should coincide with each other at the same value of $\hat{u}$. Since the characteristic length $R$ is not known, however, the time $\tau_{th}$ is not known yet. In order to find $R$ in fragile liquids, as reference one can use the dimensionless time for A particle obtained by the simulations on the SW binary mixtures \cite{toku131} or the LJ binary mixtures \cite{narumi07,narumi11} since $R$ is known as $R=\sigma$. In fact, it satisfies the power laws given by Eqs. (\ref{tau0}) and (\ref{tau}), which are described by the straight lines given in Fig. \ref{ti}(F). Then, the value of the characteristic length $R$ for particle $\alpha$ is chosen for the dimensionless time $\tau_{\beta}/\tau_{th}$ to obey those power-law lines as a function of $\hat{u}$. Similarly, in strong liquids one can use the dimensionless time for A particle of non-network glass formers A$_{80}$B$_{20}$ for $Q>Q_c$ as reference since $R$ is known as $R=\sigma$ \cite{toku131}. In fact, it satisfies the power laws given by Eqs. (\ref{tau0}) and (\ref{tau}), which are described by the straight lines given in Fig. \ref{ti}(S). In Fig. \ref{ti}, all the data for the dimensionless time $\tau_{\beta}/\tau_{th}$ are then shown versus $\hat{u}$ in each type of liquids, (F) and (S), where the fitting value of $R$ for each atom is listed in Table \ref{table-4}. Thus, there exists a small difference between fragile liquids and strong liquids. This difference would be roughly explained to result from the fact that the network strong liquids such as SiO$_2$ have an open tetrahedral network, while the fragile liquids do not \cite{side,habasaki,ediger,pas,pablo,kohira07,toku071,mei08,sal13}. The static properties of SiO$_2$ are known to be reproduced by employing the NV potential \cite{nv} and also the BKS potential \cite{woo,jac,tsune,vashi,wil,angell98,kob99,hor2,car07,voi08}. As is shown in Fig. \ref{Sq}, the static structure factor $S_{\alpha\alpha}(q)$ in network strong liquids usually has the so-called first sharp diffraction peak \cite{zallen}, while it does not in fragile liquids. On the other hand, in non-network strong liquids, their structure factors are independent of $Q$ and do not have a first sharp diffraction peak (see Fig. \ref{Sq}). However, we note here that the characteristic length $R$ of B particle increases as $Q$ increases, while that of A particle does not (see Table \ref{table-4}). This background slow motion over a wide spatial range might be a reason for the system to show the same strong properties in $f(x)$ and $\tau_i$ as those of SiO$_2$. 

In the following, we discuss three types of liquids, (F$_{non}$) non-network fragile liquids, (S$_{net}$) network strong liquids, and (S$_{non}$) non-network strong liquids, separately and then show that in each type of liquids any data coincide with each other at the same value of $\hat{u}$ \cite{toku10,toku11}. Thus, the detailed analyses suggest an existence of a master curve $H_n^{(i)}$ for the mean-$n$th displacement $M_n(t)$ given by
\begin{equation}
M_n(t)=R^nH_n^{(i)}(t/\tau_{th};\hat{u}), \label{4-4}
\end{equation}
where $i$=F$_{non}$, S$_{net}$, and S$_{non}$. Thus, all the simulation data seem to collapse onto each single master curve at the same value of $\hat{u}$, although its analytic form is not known. 

\subsection{Master curve in fragile liquids}
\begin{figure}
\begin{center}
\includegraphics[width=8.0cm]{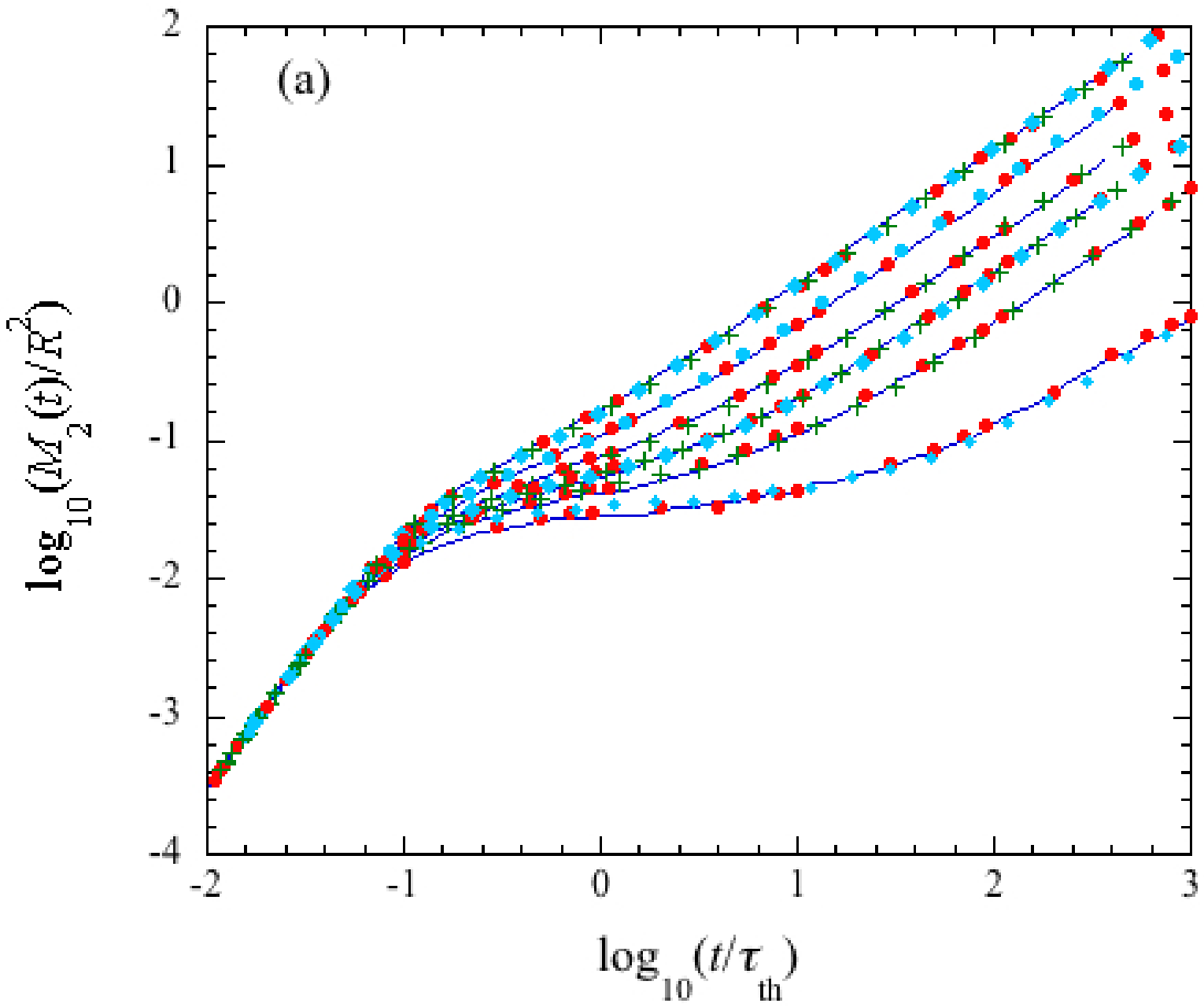}\hspace{0.0cm}
\includegraphics[width=8.0cm]{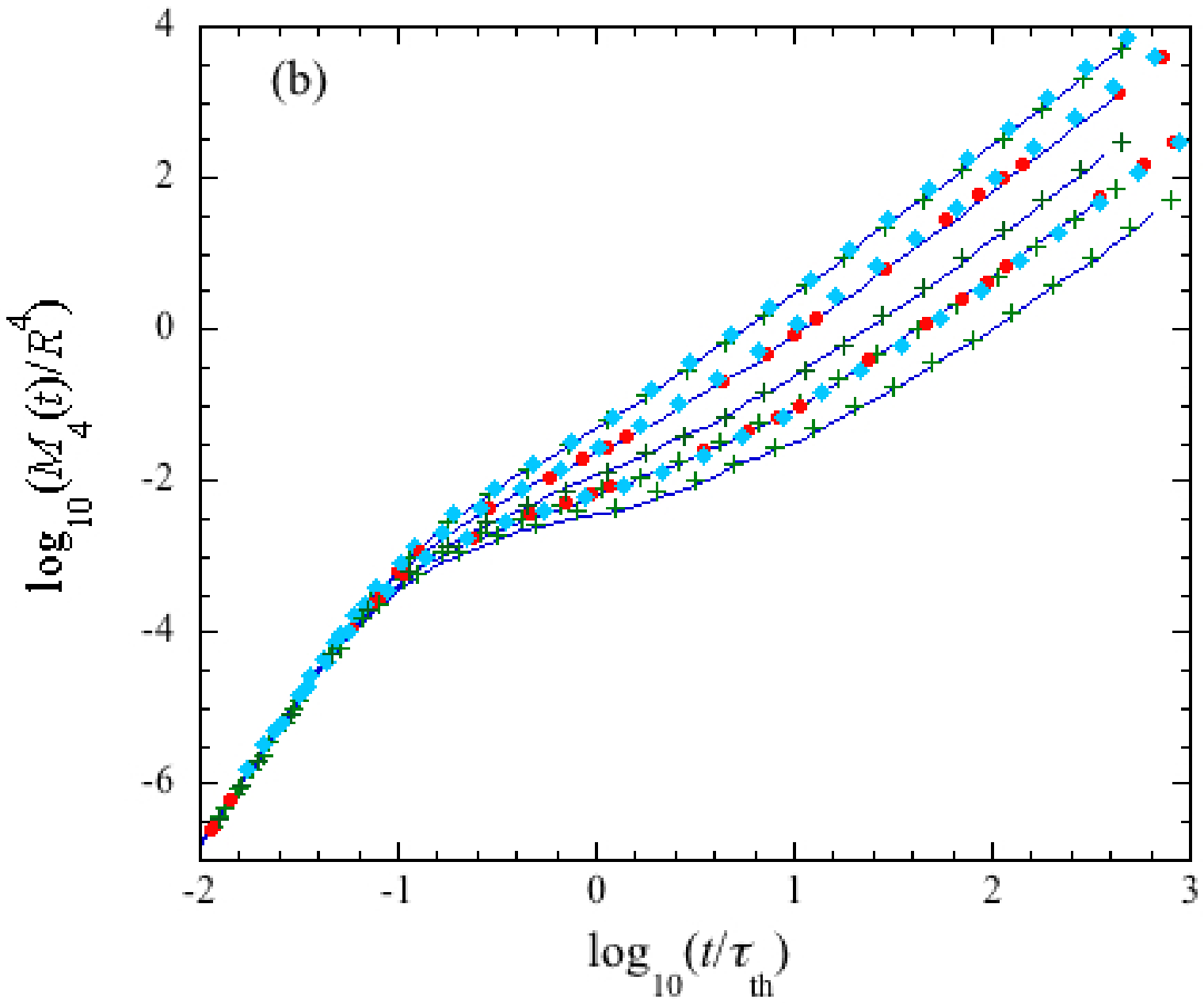}\hspace{0.0cm}
\includegraphics[width=8.0cm]{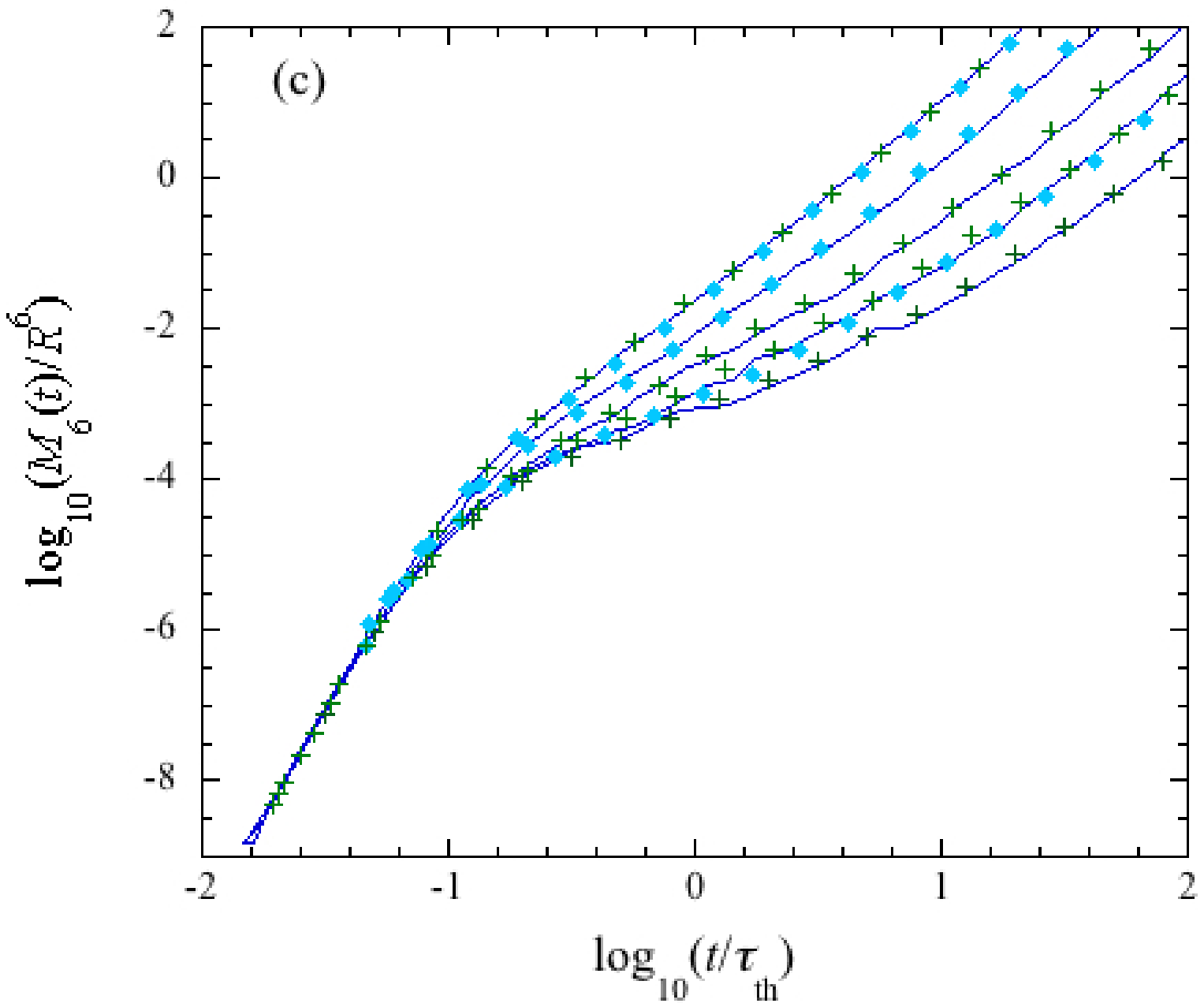}\hspace{0.0cm}
\end{center}
\caption{(Color online) A log-log plot of scaled mean-$n$th displacement $M_n(t)/R^n$ versus $t/\tau_{th}$ for different values of $\hat{u}$ in fragile liquids. (a) $n=2$ at $\hat{u}\simeq$1.677, 1.957, 2.307, 2.580, 2.926, and 3.946, and (b) $n=4$ and (c) $n=6$ at $\hat{u}\simeq$1.677, 1.957, 2.307, 2.580, and 2.926 (from top to bottom). The solid lines indicate the simulation results for Al (BM). The symbols indicate the simulation results for $(\bullet)$ A (LJ), $(\Diamond)$ A (SW $Q=1$), and $(+)$ O (BM). Temperature at each $\hat{u}$ is listed in Table \ref{table-5}.}
\label{mnf}
\end{figure}

We first show the master curve in fragile liquids. In Fig. \ref{mnf}, the simulation results for the scaled mean-$n$th displacement $M_n(t)/R^n$ are plotted versus $t/\tau_{th}$ for different values of $\hat{u}$ in fragile liquids, Al$_2$O$_3$, A$_{80}$B$_{20}$ (LJ), and A$_{80}$B$_{20}$ (SW $Q=1$), where $n=$2, 4, and 6. At each value of $\hat{u}$, the simulation results in different systems coincide with each other within error. Thus, it is suggested that there exists a master curve $H_n^{(F_{non})}(t/\tau_{th};\hat{u})$ for any fragile liquids, where the existence of $H_2^{(F_{non})}$ for many different fragile systems has already been discussed in the previous papers \cite{toku10,toku11}.
\begin{table}
\caption{Temperature versus $\hat{u}$ in fragile liquids.}
\begin{center}
\begin{tabular}{cccccc}
\hline
state & $\hat{u}$& A(LJ)  & Al(BM) &O(BM) &A(SW)$(Q=1)$\\
\hline
[L]&1.677 & 1.428 & 6000(K) & 5500(K)  & 1.667\\
&1.957 & 1.000 & 4500(K) & - & 1.250\\
&2.307& 0.769 & 3600(K) & 3400(K)  & - \\
&2.580 &0.667&3200(K)&3000(K)&0.833\\
\hline
[S]&2.926 &0.588&2800(K)&2700(K)&-\\
&3.946 &0.476& 2400(K)& -& 0.625\\
\hline
\end{tabular}
\end{center}
\label{table-5}
\end{table}

\subsection{Master curve in strong liquids}
We next show the master curve in strong liquids. We discuss two types of strong liquids, S$_{net}$ and S$_{non}$, separately.

\subsubsection{Network strong liquids}
\begin{figure}
\begin{center}
\includegraphics[width=8.0cm]{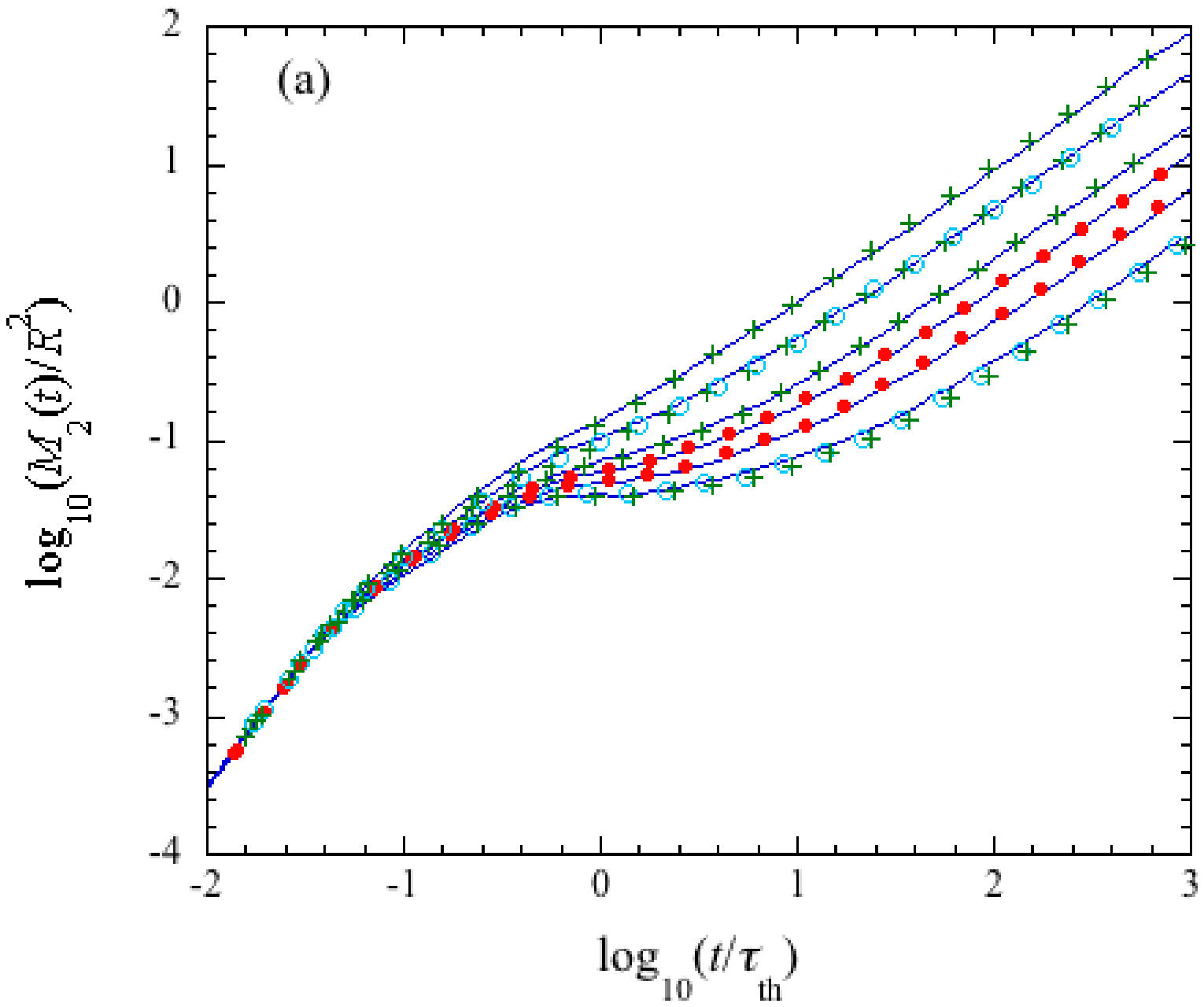}\hspace{0.0cm}
\includegraphics[width=8.0cm]{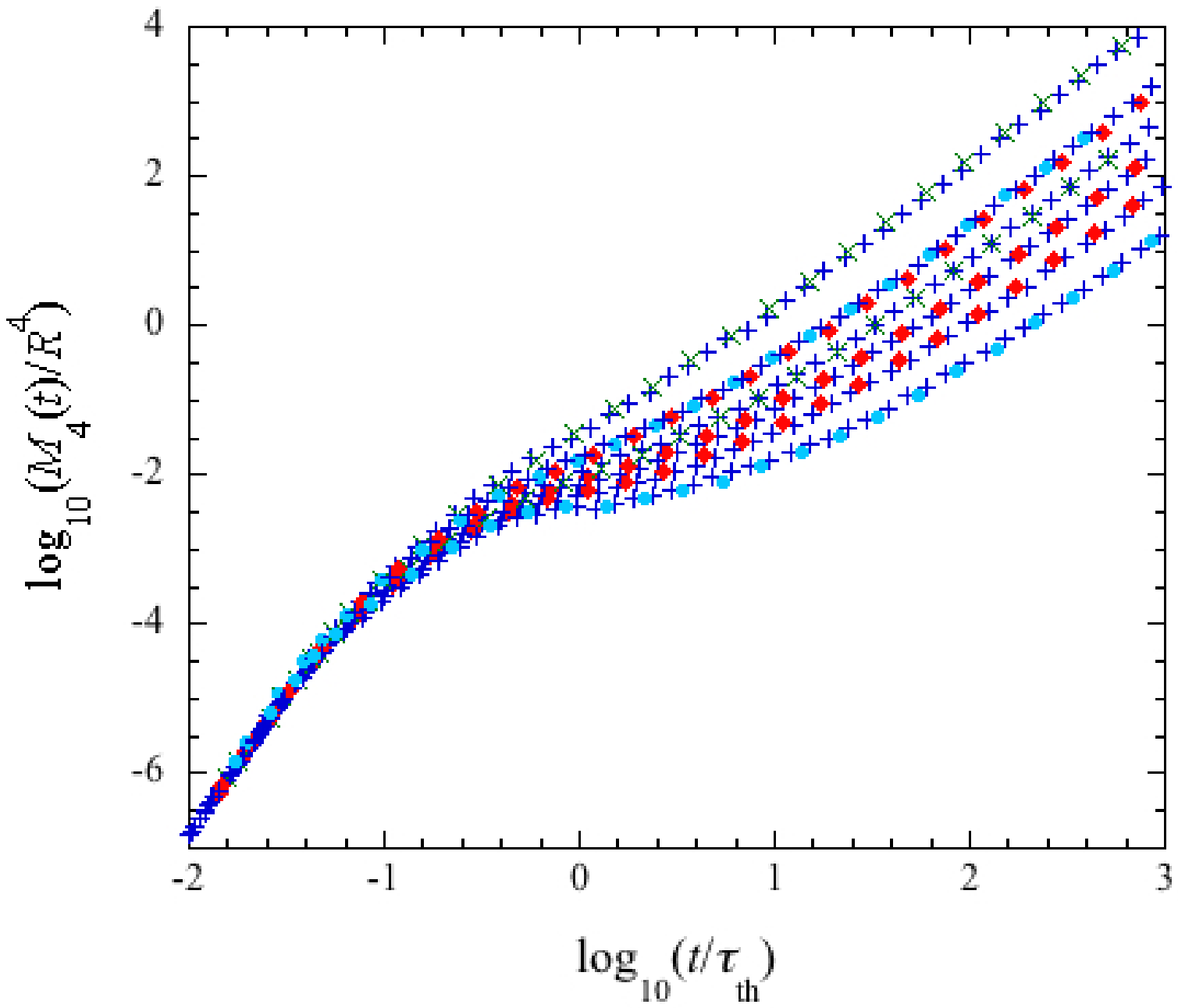}\hspace{0.0cm}
\includegraphics[width=8.0cm]{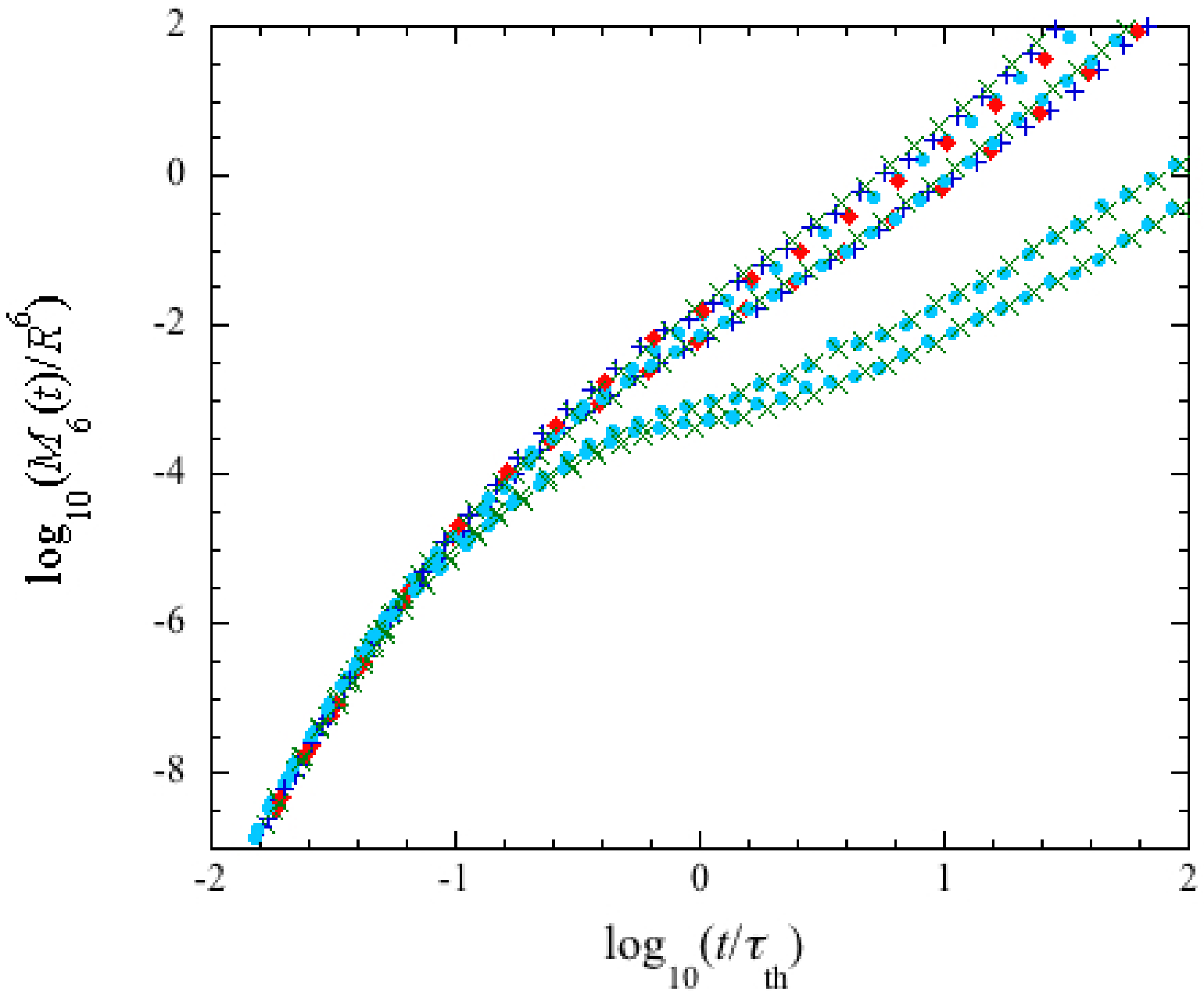}\hspace{0.0cm}
\end{center}
\caption{(Color online) A log-log plot of scaled mean-$n$th displacement $M_n(t)/R^n$ versus $t/\tau_{th}$ for different values of $\hat{u}$ in network strong liquids. (a) $n=2$ at $\hat{u}\simeq$1.793, 2.085, 2.474, 2.693, 2.955, and 3.284, (b) $n=4$ at $\hat{u}\simeq$1.793, 2.085, 2.474, 2.693, 2.955, and 3.284, and (c) $n=6$ at $\hat{u}\simeq$1.793, 1.913, 2.085, 3.021, 3.322 (from top to bottom). The solid lines indicate the simulation results for Si (NV). The symbols indicate the simulation results for $(\bullet)$ O(NV), $(+)$ O(BKS), and $(\circ)$ Si(BKS). Temperature at each $\hat{u}$ is listed in Table \ref{table-6}.}
\label{mns}
\end{figure}

We first discuss network strong liquids whose static structure factor has a first sharp diffraction peak (see Fig. \ref{Sq}). In Fig. \ref{mns}, the simulation results for the mean-$n$th displacement $M_n(t)/R^n$ are plotted versus $t/\tau_{th}$ for different values of $\hat{u}$ in network glass formers, SiO$_2$ (BKS) and SiO$_2$ (NV), where $n=$2, 4, and 6. At each value of $\hat{u}$, the simulation results in different systems coincide with each other within error. Thus, this suggests an existence of a single master curve $H_n^{(S_{net})}(t/\tau_{th};\hat{u})$ for any network glass formers.
\begin{table}
\caption{Temperature versus $\hat{u}$ in network strong liquids.}
\begin{center}
\begin{tabular}{cccccc}
\hline
state & $\hat{u}$& Si(NV)& O(NV)&Si(BKS)&O(BKS)\\
\hline
[L]&1.793 &5000(K) & - & -&5800(K)\\
&1.913 &-&5000(K)& 5200(K)&-\\
&2.085 & 4400(K) & - & 4800(K)  & 5000(K)\\
&2.220&4200(K) & 4300(K) & 4600(K) &-\\
&2.474 & 3900(K) & - & - & 4400(K)\\
&2.693 &3700(K)&3800(K)&-&-\\
\hline
[S]&2.955 &3500(K)& 3600(K)& -& -\\
&3.021&-&-&3800(K)&3800(K)\\
&3.282& 3300(K) & - & 3600(K)  & 3600(K) \\
&3.322& -&-&3600(K)&3600(K)\\
\hline
\end{tabular}
\end{center}
\label{table-6}
\end{table}

\subsubsection{Non-network strong liquids}
\begin{figure}
\begin{center}
\includegraphics[width=8.0cm]{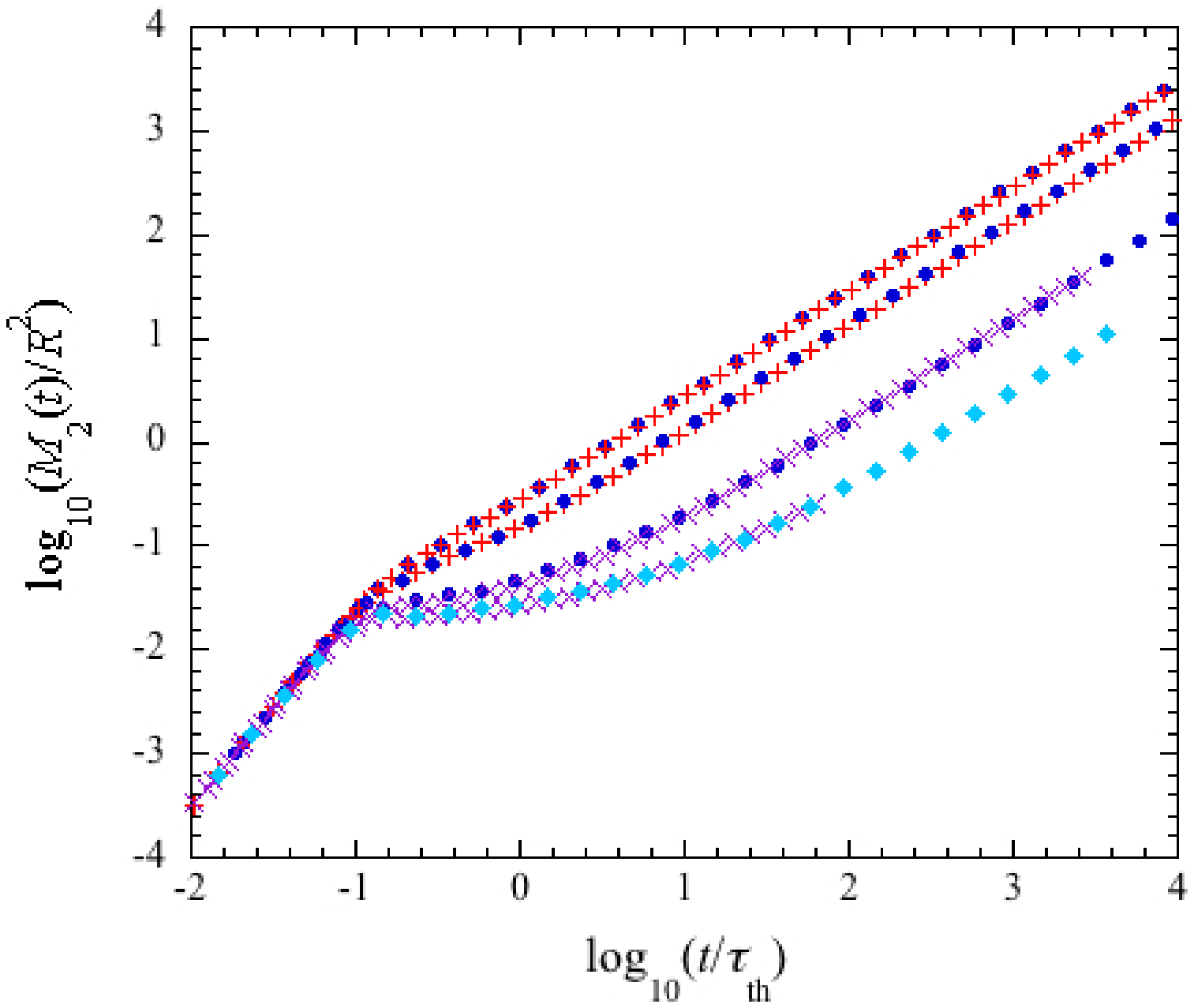}\hspace{0.0cm}
\includegraphics[width=8.0cm]{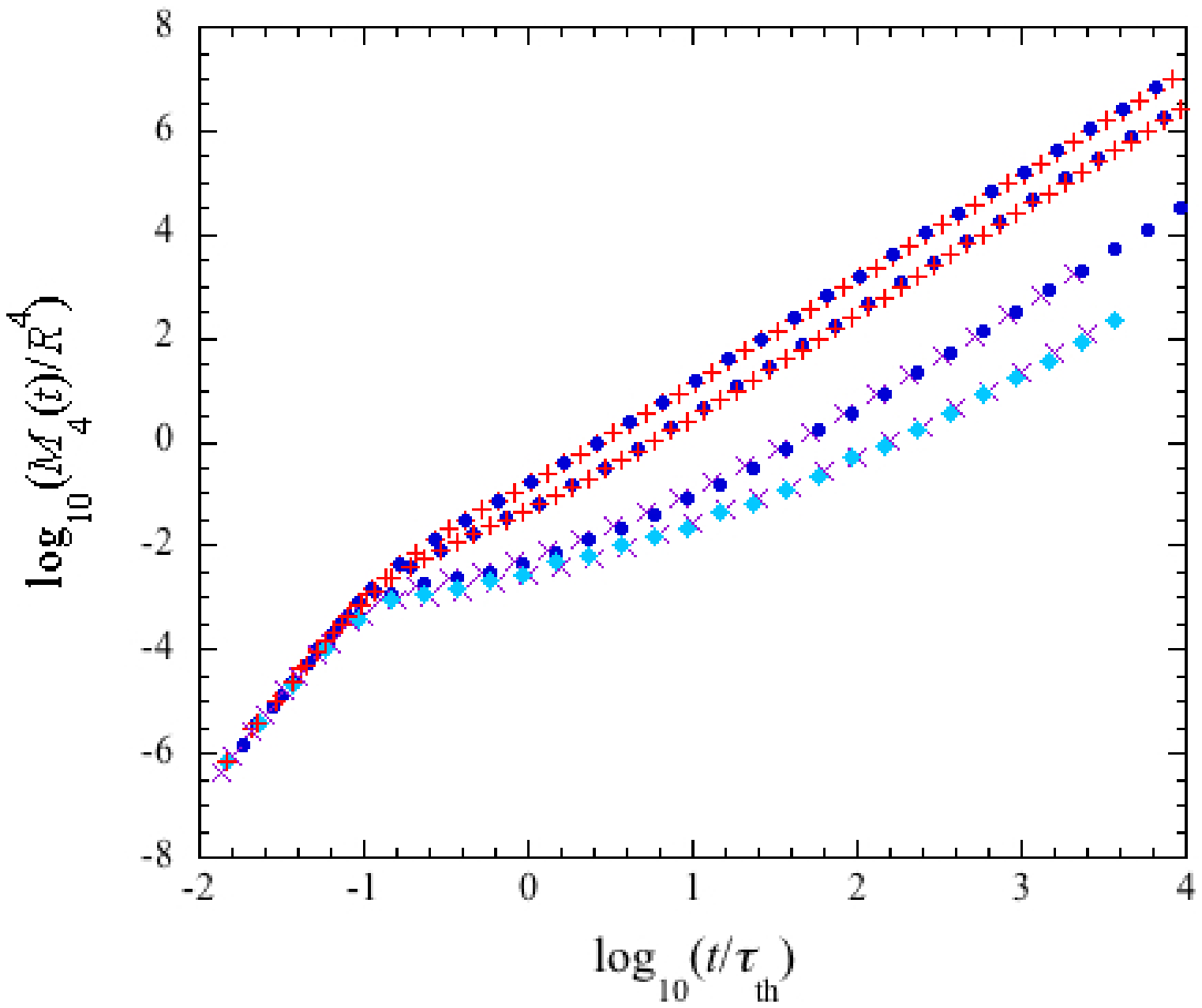}\hspace{0.0cm}
\includegraphics[width=8.0cm]{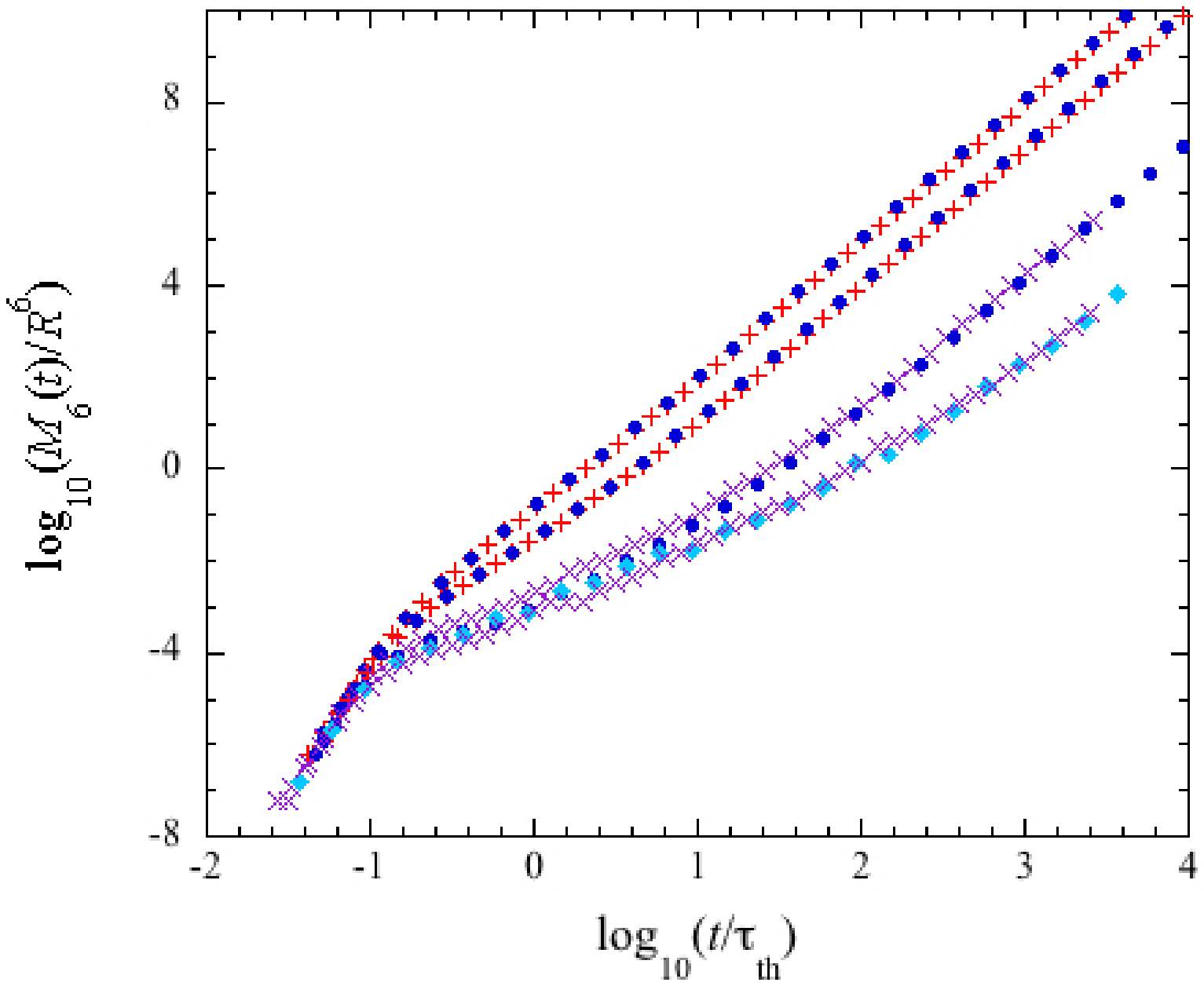}\hspace{0.0cm}
\end{center}
\caption{(Color online) A log-log plot of scaled mean-square displacement $M_n(t)/R^n$ versus dimensionless time $t/\tau_{th}$ for different values of $\hat{u}$ in non-network strong liquids. (a) $n=2$, (b) $n=4$, and (c) $n=6$ at $\hat{u}\simeq$1.328, 1.667, 2.602, and 3.257  from top to bottom. The symbols indicate the simulation results; $(+)$ A($Q=100$), $(\times)$ B($Q=100$), $(\bullet)$ A($Q=50$), and $(\Diamond)$ B($Q=50$). Temperature at each $\hat{u}$ is listed in Table \ref{table-7}.}
\label{nsn}
\end{figure}

We next discuss non-network strong liquids whose static structure factor has no first sharp diffraction peak. In Fig. \ref{nsn}, the simulation results for the mean-$n$th displacement $M_n(t)/R^n$ are plotted versus $t/\tau_{th}$ for different values of $\hat{u}$ in non-network glass formers, A$_{80}$B$_{20}$ (SW $Q=50$) and A$_{80}$B$_{20}$ (SW $Q=100$), where $n=$2, 4, and 6. At each value of $\hat{u}$, the simulation results in different systems coincide with each other within error. Thus, this also suggests an existence of a master curve $H_n^{(S_{non})}(t/\tau_{th};\hat{u})$ for any non-network glass formers.
\begin{table}
\caption{Temperature versus $\hat{u}$ in non-network strong liquids.}
\begin{center}
\begin{tabular}{cccccc}
\hline
state & $\hat{u}$& A($Q=100$)& B($Q=100$) &A($Q=50$)&B($Q=50$)\\
\hline
[L]&1.328 &5.000 & - & 5.000&-\\
&1.667 &2.500&2.500& -&-\\
&2.602 & - & 0.714 & 1.000 & -\\
\hline
[S]&3.257 &-&0.625&-&0.625\\
\hline
\end{tabular}
\end{center}
\label{table-7}
\end{table}

\begin{figure}
\begin{center}
\includegraphics[width=8.0cm]{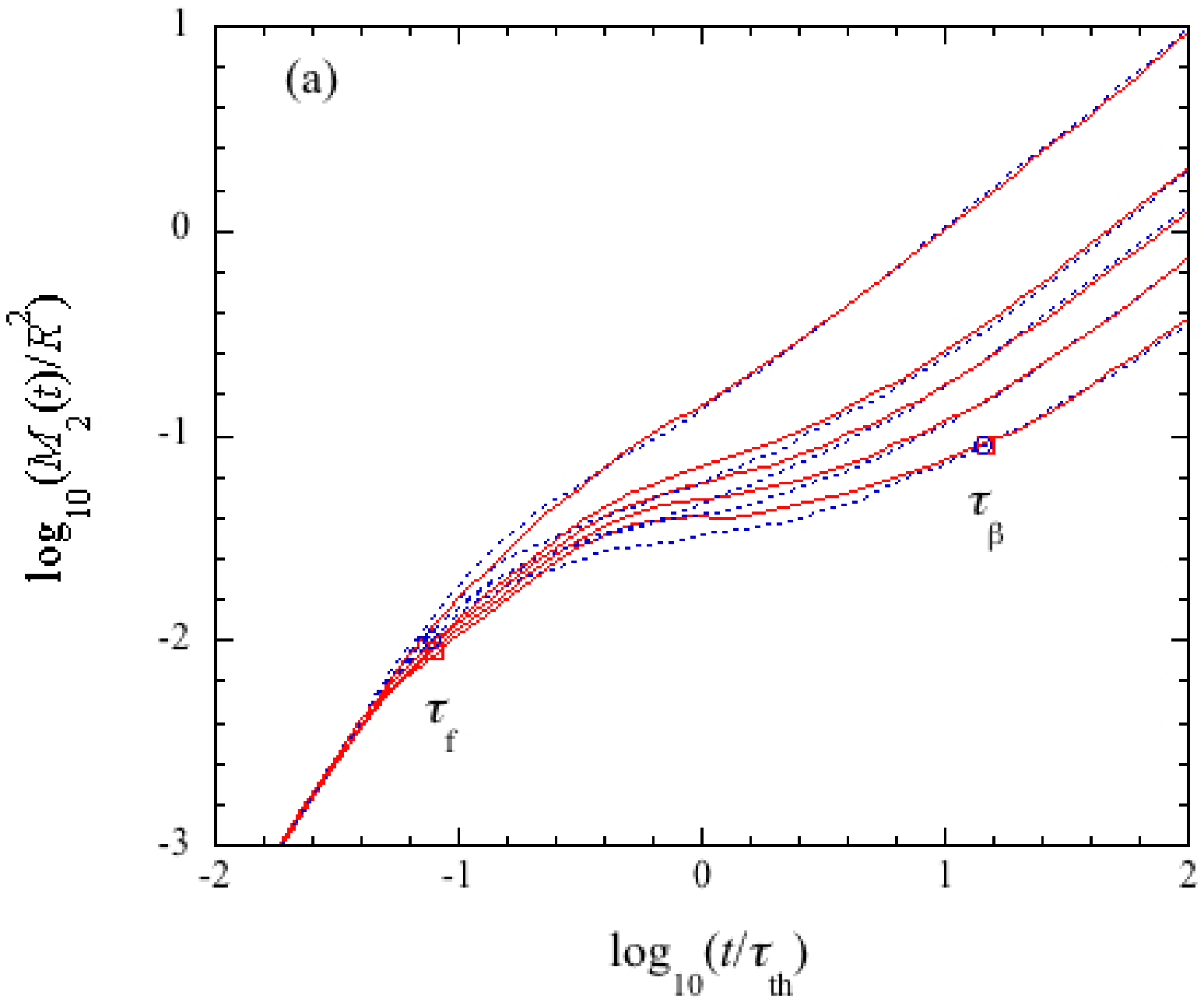}\hspace{0.0cm}
\includegraphics[width=8.0cm]{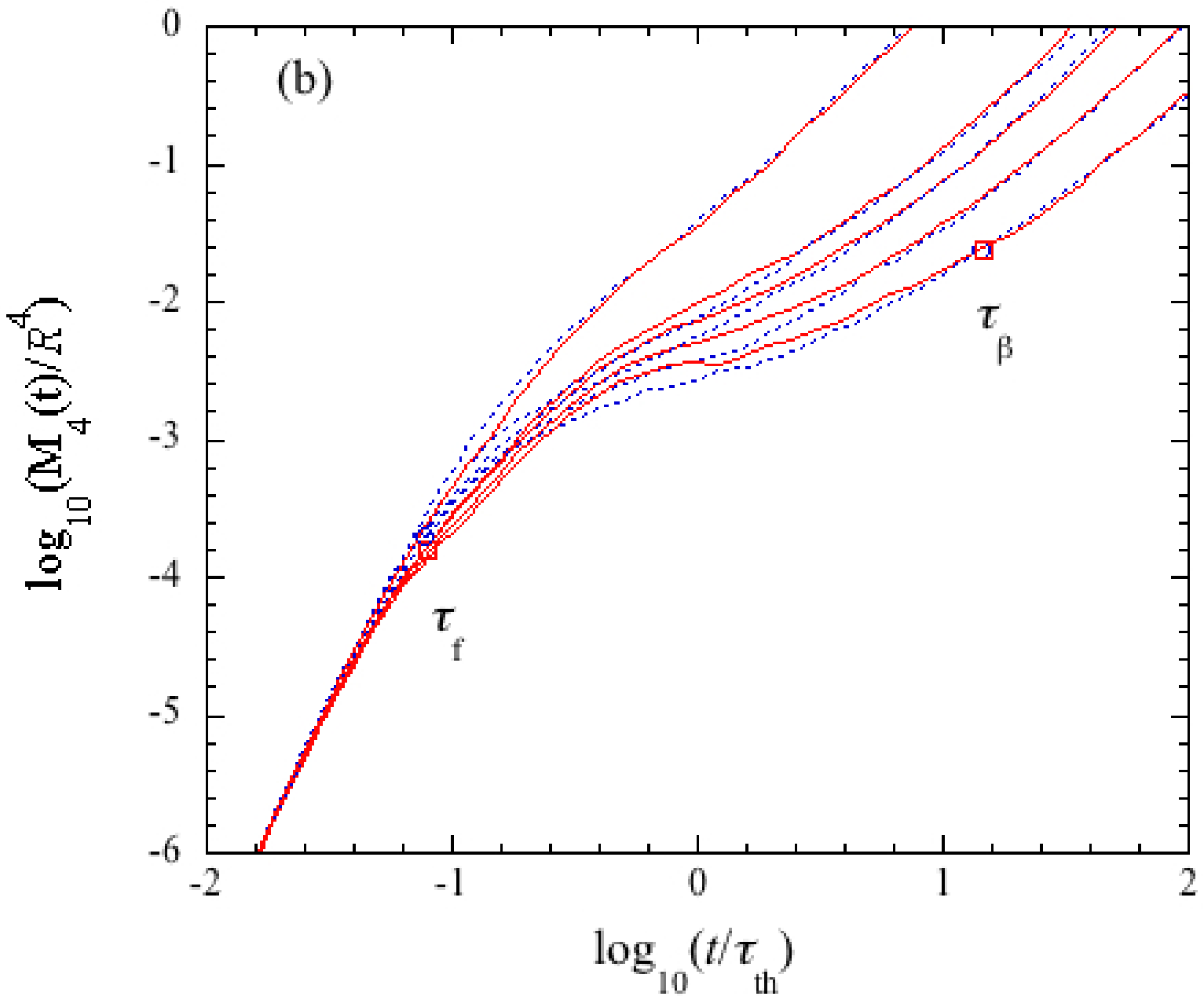}\hspace{0.0cm}
\includegraphics[width=8.0cm]{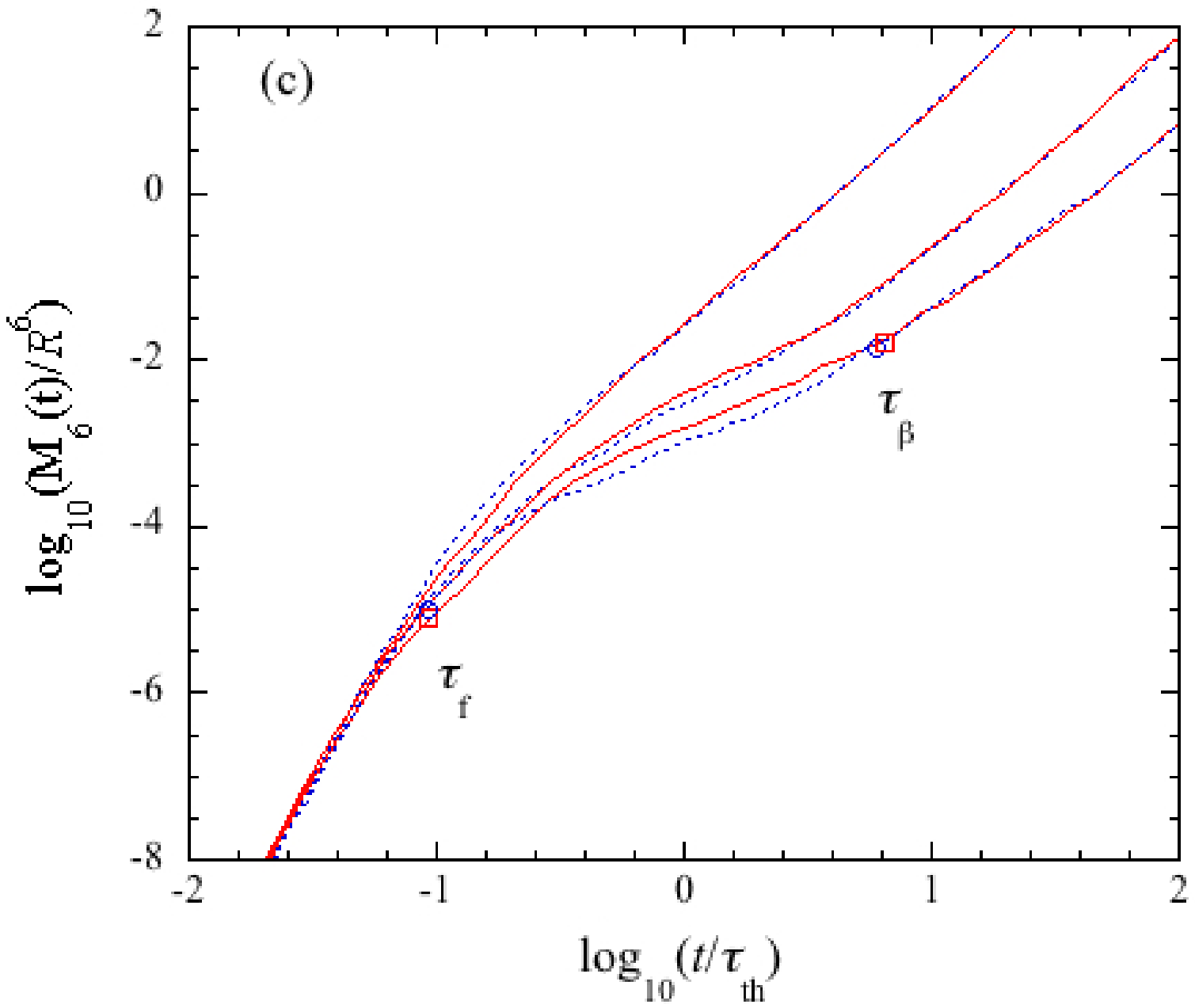}\hspace{0.0cm}
\end{center}
\caption{(Color online) A log-log plot of scaled mean-$n$th displacement $M_n(t)/R^n$ versus $t/\tau_{th}$ for different value of $\hat{u}$ in fragile liquids Al$_2$O$_3$ and network strong liquids SiO$_2$. (a) $n=2$ and (b) $n=4$ at $\hat{u}\simeq$1.779, 2.494, 2.638, 2.949, and 3.274, and (c) $n=6$ at $\hat{u}\simeq$1.669, 2.369, and 2.745 (from top to bottom). The solid lines indicate the simulation results for (a) Si (NV), (b) Si (NV), and (c) Si (BKS). The dotted lines indicate the simulation results for Al. The symbols indicate the relaxation times $\tau_f$ and $\tau_{\beta}$ at $\hat{u}=3.274$ for (a) and (b) and 2.745 for (c); $(\Box)$ for Si and $(\circ)$ for Al. Temperature at each $\hat{u}$ is listed in Table \ref{table-8}.}
\label{mnfsAl}
\end{figure}
\section{Differences between fragile liquids and strong liquids}
In the previous section, we have shown that there exists a master curve $H_n^{(i)}(t/\tau_{th};\hat{u})$ in each type of liquids. In the present section, therefore, we compare the dynamics of strong liquids with that of fragile liquids at the same value of $\hat{u}$ and explore how their dynamics is different from each other.

\subsection{Network strong liquids versus fragile liquids}
We first compare the simulation results for the network strong liquids with those for fragile liquids at the same value of $\hat{u}$. In Fig. \ref{mnfsAl}, the simulation results for the mean-$n$th displacement $M_n(t)/R^n$ are plotted versus $t/\tau_{th}$ for different values of $\hat{u}$ in fragile liquids Al$_2$O$_3$ (BM) and network glass formers SiO$_2$ (NV, BKS), where $n=$2, 4, and 6. As a typical example, the dynamics of Si is compared with that of Al. In the $\beta$ stage for $\tau_f\leq t\leq \tau_{\beta}$, the dynamical behavior of Si is quite different from that of Al. This difference is caused by an open tetrahedral network in SiO$_2$. The results obtained here are also seen for other combinations between Al$_2$O$_3$ and SiO$_2$. 
\begin{table}
\caption{Temperature versus $\hat{u}$ in fragile liquids and strong liquids.}
\begin{center}
\begin{tabular}{ccccccc}
\hline
state & $\hat{u}$& Al(BM)& Si(NV)&Si(BKS)&A($Q=100$)\\
\hline
[L]&1.779 &5300(K)& 5000(K)&-&-\\
&1.842 &5000(K)& -&-&2.000\\
&1.969 &4500(K)& -&-&1.667\\
&2.302 &3600(K)& -&-&1.250\\
&2.369&3500(K)&-&4400(K)&-\\
&2.494 & 3300(K) & 3900(K) & -&-\\
&2.638 & 3100(K)& 3700(K) & -  & 1.000\\
\hline
[S]&2.745&3000(K)&-&4000(K)&-\\
&2.949 &2800(K)&3500(K)&-&-\\
&3.143 & 2700(K) & 3400(K)& -& 0.833\\
&3.274 & 2600(K) & 3300(K)  & -&-\\
\hline
\end{tabular}
\end{center}
\label{table-8}
\end{table}

\subsection{Non-network strong liquids versus fragile liquids}
\begin{figure}
\begin{center}
\includegraphics[width=8.0cm]{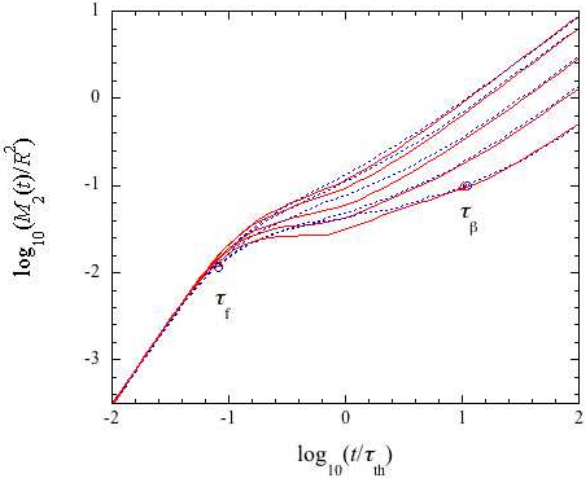}
\end{center}
\caption{(Color online) A log-log plot of scaled mean-square displacement $M_2(t)/R^2$ versus $t/\tau_{th}$ for different value of $\hat{u}$ in fragile liquids Al$_2$O$_3$ and non-network strong liquids A$_{80}$B$_{20}$ (SW $Q=100$), where $\hat{u}\simeq$1.842, 1.969, 2.302, 2.638, and 3.143. The dotted lines indicate the simulation results for Al and the solid lines for A ($Q=100$). The symbols indicate the relaxation times $\tau_f$ and $\tau_{\beta}$ at $\hat{u}=3.143$; $(\triangle)$ for A ($Q=100$) and $(\circ)$ for Al. Temperature at each $\hat{u}$ is listed in Table \ref{table-8}.}
\label{AQAl}
\end{figure}
We next compare the simulation results for the non-network strong liquids with those for fragile liquids at the same value of $\hat{u}$. In Fig. \ref{AQAl}, the simulation results for the mean-square displacement $M_2(t)/R^2$ are plotted versus $t/\tau_{th}$ for different values of $\hat{u}$ in fragile liquids Al$_2$O$_3$ (BM) and non-network strong liquids A$_{80}$B$_{20}$ (SW $Q=100$). The dynamics of A is compared with that of Al. In the $\beta$ stage for $\tau_f\leq t\leq \tau_{\beta}$, the dynamical behavior of A ($Q=100$) is shown to be quite different from that of Al. This must be caused by the fact that A particle moves through the interactions with slowly-moving B particles. The results obtained here are also seen for other combinations between Al$_2$O$_3$ and A$_{80}$B$_{20}$ in $M_n(t)$.

\begin{figure}
\begin{center}
\includegraphics[width=8.0cm]{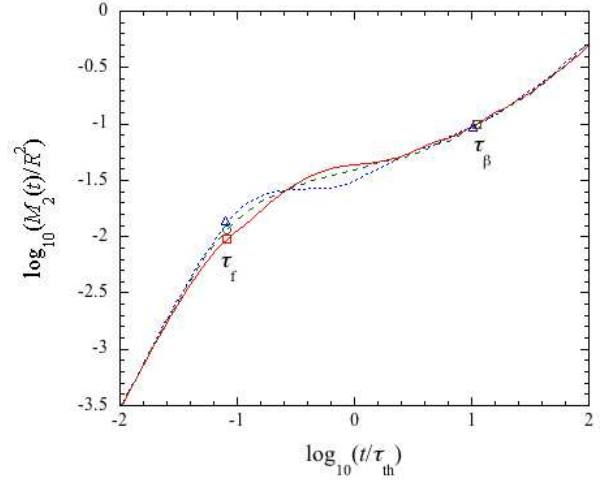}
\end{center}
\caption{(Color online) A log-log plot of scaled mean-square displacement $M_2(t)/R^2$ versus $t/\tau_{th}$ at $\hat{u}\simeq$3.143 for network strong liquids SiO$_2$ (NV), non-network strong liquids A$_{80}$B$_{20}$ (SW $Q=100$), and non-network fragile liquids Al$_2$O$_3$ (BM). The solid lines indicate the simulation results for Si at $T=3400$ (K), the dotted lines for A ($Q=100$) at $T=0.833$, and the dashed line for Al at $T=2700$ (K). The symbols indicate the relaxation times $\tau_f$ and $\tau_{\beta}$; ($\Box$) for Si, ($\triangle$) for A ($Q=100$), and ($\circ$) for Al.}
\label{m2fss}
\end{figure}
\subsection{Network strong liquids versus non-network strong liquids}
In the $\beta$ stage, the dynamical behavior of both network strong liquids and non-network strong liquids have been shown to be quite different from that of fragile liquids at the same value of $\hat{u}$. Hence we now compare the dynamics of network strong liquids with that of non-network strong liquids. In Fig. \ref{m2fss}, the simulation results for the mean-square displacement $M_2(t)/R^2$ are plotted versus $t/\tau_{th}$ for network glass formers SiO$_2$ (NV) and non-network glass formers A$_{80}$B$_{20}$ (SW $Q=100$). As a typical example, the value of $\hat{u}$ is chosen in a supercooled state as $\hat{u}\simeq 3.143$ because the difference between both glass formers is seen clearly. In the $\beta$ stage for $\tau_f\leq t\leq \tau_{\beta}$, the dynamics of Si is shown to be different from that of A ($Q=100$), although both glass formers obey the strong master curve given by $f(x;\eta=5/3)$. This difference results from the fact that SiO$_2$ has a network structure, while A$_{80}$B$_{20}$ does not.  The results obtained here are also seen for other combinations between SiO$_2$ and A$_{80}$B$_{20}$ (SW $Q>20$) at any temperatures. For comparison, the simulation results for Al are also plotted. We note here that A particle moves faster than Al around $\tau_f$. This means that A particle is weakly influenced by the correlation of B particle because B particle hardly moves on a time scale of $\tau_f$.

\section{Summary}
In the present paper, we have proposed the systematic method to investigate how the dynamics of strong liquids is different from that of fragile liquids. As examples of glass-forming materials, we have taken Al$_2$O$_3$, the LJ binary mixture A$_{80}$B$_{20}$, the SW binary mixture A$_{80}$B$_{20}$, and SiO$_2$ (BKS and NV). We have first applied the mean-field theory for the simulation results in those different glass-forming materials. Then, we have obtained the long-time self-diffusion coefficient $D(T)$ and also the mean-field values of the characteristic time $\tau_{\beta}$. By using the master curve $f(x;\eta)$, the diffusion coefficients in different systems have been safely classified into two types of liquids, fragile liquids and strong liquids. Thus,  Al$_2$O$_3$, the LJ binary mixture A$_{80}$B$_{20}$, and the SW binary mixture A$_{80}$B$_{20}$ with $Q<Q_c$ were classified as fragile liquids with $\eta=4/3$, while SiO$_2$ (BKS and NV) and the SW binary mixture A$_{80}$B$_{20}$ with $Q>Q_c$ were classified as strong liquids with $\eta=5/3$ (see Fig. \ref{d}). In order to use the universality that all the dimensionless physical quantities must coincide with each other at the same value of the universal parameter $\hat{u}(=\log_{10}(Rv_{th}/D))$, we have then adjusted the unknown characteristic length $R$ in each system so that the dimensionless time $\tau_{\beta}/\tau_{th}$ coincides with that obtained by the simulations on the SW binary mixture since $R$ is known to be $\sigma$ there. The results are shown in Fig. \ref{ti} and Table \ref{table-4}. We have next investigated the simulation results for $M_n(t)$ in each type, (F) and (S), at the same value of $\hat{u}$. In type (F$_{non}$), all the simulation results have been shown to collapse onto a single master curve $H_n^{(F_{non})}$ at the same value of $\hat{u}$ (see Figs. \ref{mnf}). On the other hand, in type (S) two different master curves, $H_n^{(S_{net})}$ and $H_n^{(S_{non})}$, have been shown to exist, depending on whether the static structure factor has the so-called first sharp diffraction peak or not. The master curve $H_n^{(S_{net})}$ stands for network glass formers (S$_{net}$), such as SiO$_2$, while $H_n^{(S_{non})}$ for non-network glass formers (S$_{non}$), such as A$_{80}$B$_{20}$ with $Q>Q_c$, whose static structure factor has structural properties similar to those of (F). Thus, all the simulation results in each type have been shown to collapse onto each master curve at the same value of $\hat{u}$ (see Figs. \ref{mns} and \ref{nsn}). Those classifications have been done based on the fact that the simulation results of type $i$ do not coincide with those of different type $j$ in the cage region even at the same value of $\hat{u}$ (see Figs. \ref{mnfsAl}, \ref{AQAl}, and \ref{m2fss}). In fact, the disagreement between (F$_{non}$) and (S$_{net}$) is reasonable because their static structure factors have different structural properties from each other. On the other hand, the static structure factor of (S$_{non}$) has structural properties similar with that of (F$_{non}$). In (S$_{non}$), however, mass of particle B is much larger than that of A particle. Hence slow dynamics occurs anti-symmetrically between A and B since the length scale $R$ of B increases as $Q$ increases, while that of A is unchanged. This might cause a strong character, which leads to a difference between (F$_{non}$) and (S$_{non}$) in the cage region. This situation would be rather similar to that seen in polymer gels \cite{toku13,seki}. Finally, from a unified point of view proposed in this paper one should also investigate the other interesting glass-forming materials such as Se which has a network structure but is usually believed to be a fragile liquid (F$_{net}$). This will be discussed elsewhere.

Acknowledgments

Authors (M.T. and J.K.) wish to thank Prof. Takashi Nakamura for his hospitality and encouragement. This work was partially supported by High Efficiency Rare Elements Extraction Technology Area, Institute of Multidisciplinary Research for Advanced Materials (IMRAM), Tohoku University, Japan. The simulations were performed by using the SGI Altix3700Bx2 in Advanced Fluid Information Research Center, Institute of Fluid Science, Tohoku University.

\end{document}